\documentclass[aps,amssymb,amsmath,preprint,floatfix,superscriptaddress]{revtex4-1}
\usepackage{graphicx,color}
\usepackage{epstopdf}
\usepackage{amsmath}
\if0
\usepackage{lineno}
\newcommand*\patchAmsMathEnvironmentForLineno[1]{
  \expandafter\let\csname old#1\expandafter\endcsname\csname #1\endcsname
  \expandafter\let\csname oldend#1\expandafter\endcsname\csname end#1\endcsname
  \renewenvironment{#1}
     {\linenomath\csname old#1\endcsname}
     {\csname oldend#1\endcsname\endlinenomath}}
\newcommand*\patchBothAmsMathEnvironmentsForLineno[1]{
  \patchAmsMathEnvironmentForLineno{#1}
  \patchAmsMathEnvironmentForLineno{#1*}}
\AtBeginDocument{
\patchBothAmsMathEnvironmentsForLineno{equation}
\patchBothAmsMathEnvironmentsForLineno{align}
\patchBothAmsMathEnvironmentsForLineno{flalign}
\patchBothAmsMathEnvironmentsForLineno{alignat}
\patchBothAmsMathEnvironmentsForLineno{gather}
\patchBothAmsMathEnvironmentsForLineno{multline}
}
\linenumbers
\fi
\usepackage{booktabs}
\usepackage{comment}
\usepackage{soul}
\usepackage{ulem}
\newif\iffigure
\figuretrue

\draft % marks overfull lines with a black rule on the right

\begin{document}

\title{Interface-induced field-like optical spin torque in a ferromagnet/heavy metal heterostructure}

\author{Satoshi Iihama}
\email{satoshi.iihama.d6@tohoku.ac.jp}
\affiliation{Frontier Research Institute for Interdisciplinary Sciences (FRIS), Tohoku University, Sendai 980-8578, Japan}
\affiliation{Center for Spintronics Research Network (CSRN), Tohoku University, Sendai 980-8577, Japan}

\author{Kazuaki Ishibashi}
\affiliation{Department of Applied Physics, Graduate School of Engineering, Tohoku University, 6-6-05, Aoba-yama, Sendai 980-8579, Japan}
\affiliation{WPI Advanced Institute for Materials Research (AIMR), Tohoku University, 2-1-1, Katahira, Sendai 980-8577, Japan}

\author{Shigemi Mizukami}
\affiliation{Center for Spintronics Research Network (CSRN), Tohoku University, Sendai 980-8577, Japan} 
\affiliation{WPI Advanced Institute for Materials Research (AIMR), Tohoku University, 2-1-1, Katahira, Sendai 980-8577, Japan}
\affiliation{Center for Science and Innovation in Spintronics (CSIS), Core Research Cluster (CRC), Tohoku University, Sendai 980-8577, Japan}

\date{\today}

\begin{abstract}
The manipulation of magnetization in a metallic ferromagnet by using optical helicity has been much attracted attention for future opto-spintronic devices.
The optical helicity induced torques on the magnetization, {\it optical spin torque}, have been observed in ferromagnetic thin films recently.
However, the interfacial effect of the optical spin torque in ferromagnet/nonmagnetic heavy metal heterostructures have not been addressed so far, which are widely utilized to efficiently control magnetization via electrical means.
Here, we studied optical spin torque vectors in the ferromagnet/nonmagnetic heavy metal heterostructures and observed that in-plane field-like optical spin torque was significantly increased with decreasing ferromagnetic layer thicknesses. 
The interfacial field-like optical spin torque was explained by the optical Rashba-Edelstein effect caused by the structural inversion symmetry breaking. 
This work will aid in the efficient optical manipulation of thin film nanomagnets using optical helicity.
\end{abstract}
\maketitle

\section{Introduction} 

Recently, ferromagnetic/nonmagnetic heavy metal heterostructures have been widely used for manipulating magnetization via electrical means in the field of spin-orbitronics \cite{Manchon2015}.
The spin Hall effect and/or the Rashba-Edelstein effect have been observed in heterostructures, which enable the efficient control of thin film magnets \cite{Miron_Nature, Liu_Science}.
A previous study reported that the magnetization of an ultrathin Co film sandwiched between Pt and AlO$_x$ layers can be switched by in-plane current injection. This is attributed to the torque generated by the Rashba field due to the asymmetric heterostructure \cite{Miron_Nature}. 
Further, it was demonstrated that the spin-transfer torque in a ferromagnetic layer can be induced by the spin Hall effect in adjacent nonmagnetic heavy metal layers such as Ta and Pt \cite{Liu_Science}. 
Both these torques, which are generally called spin-orbit torques, can be considered to be caused by the interface effect of the heterostructure and the bulk effect of the nonmagnetic layer \cite{Haney2013, Amin2016}.

Analogous to the electrical generation of the spin-orbit torque, the optical manipulation of magnetization by using the spin-angular momentum of light, namely optical helicity, in metallic thin films has been investigated extensively in experimental \cite{Lambert2014, Hadri2016, Takahashi2016, John2017, Kichin2019} and theoretical studies \cite{Battiato2014, Berritta2016, Freimuth2016}. 
This optical manipulation will have future applications in ultrafast and low energy consumption opto-spintronic devices \cite{Kimel2019, Becker2020}.
Although the transfer of spin angular momentum from light to thin film metallic magnets has been considered to be marginal, accumulative magnetization switching induced by optical helicity has been observed \cite{Lambert2014, Hadri2016, Takahashi2016, John2017, Kichin2019}, which stems from magnetic circular dichroism \cite{Ellis2016, Gorchon2016}.
In contrast to the accumulative magnetization switching induced by magnetic circular dichroism, a recent experimental study observed that magnetization reversal is also induced by a single circularly polarized laser pulse \cite{Kichin2019}.
In addition, magnetization precession dynamics induced by optical helicity in a metallic thin film magnet was observed \cite{Choi2017, Choi2020}.
The torque on magnetization induced by optical helicity in metallic heterostructures can generate a helicity-dependent terahertz emission \cite{Huisman2016, Jungfleish2018}, in which the terahertz photocurrent is produced due to the inversion symmetry breaking of the thin film structure.
These experiments indicated that the optical helicity induced torque, which we term {\it optical spin torque}, has a certain effect on metallic thin film magnets, although many previous studies have used magnetic semiconductors or insulators to observe the effect of optical helicity \cite{Lampel1967, Awschalom1986, Oiwa2002, Kimel2005, Satoh2010, Nemec2012, Tesarova2013, Ramsay2015}.
\color{black}
Whereas optical spin transfer torque as well as optical spin orbit torque have been observed in a magnetic semiconductor GaMnAs, respectively in Ref. \cite{Nemec2012} and Ref. \cite{Tesarova2013}, the mechanism of the torque is different with the case of the metal.
\color{black}

In the initial observations of optical spin torque in metallic thin film magnets, optical helicity induced magnetization precession was mainly explained by the inverse Faraday effect in ferromagnetic metals \cite{Choi2017}, in which the optical spin torque has a field-like form, ${\bf m} \times {\bf s}$, where ${\bf m}$ and ${\bf s}$ are the magnetization and spin direction pointing along the wave vector of the light, i.e., ${\bf s} \propto {\bf E} \times {\bf E}^{\star }$.
However, the recent observations of the optical helicity induced magnetization precession in Co/Pt bilayer can be explained using spin-transfer mechanism as follows \cite{Choi2020}. 
The spin ${\bf s}$ generated by the optical helicity in the Pt layer was due to the optical orientation, and it transfers its angular momentum to an adjacent Co layer via the spin-transfer torque effect. 
It was concluded that the magnetization precession in Co was excited by the out-of-plane spin-transfer torque in the form ${\bf m}\times ({\bf m} \times {\bf s})$ as shown in Fig. 1(a).

\color{black}
Ref. \cite{Choi2020} have studied the effect of circularly polarized light with different thickness of Co and Pt. 
Then, they discussed the thickness dependences in terms of light absorption and optical orientation in non-magnetic heavy metal Pt. 
However, such spin-orbit induced optical effect can also arise from interfaces, like optical Rashba-Edelstein effect proposed in the past\cite{Edelstein1997}.
In this article, we experimentally study the optical spin torque for magnetic and heavy metal heterostructures in detail and report, for the first time, a new type of {\it interface-induced} field-like optical spin torque, as schematically shown in Fig. 1(b).
\color{black}

\section{Experimental procedure}

Thin film samples were prepared by the magnetron sputtering method. 
MgO (10)/Fe$_{50}$Co$_{50}$ ($d_{\rm FeCo}$)/Pt ($d_{\rm Pt}$) (thickness is in nm) films were deposited on thermally oxidized Si/SiO$_2$ substrate. 
To measure circularly polarized laser pulse induced magnetization dynamics, the conventional time-resolved magneto-optical Kerr effect (MOKE) setup was employed \cite{Iihama2014, Mizukami2016}.
The wavelength, pulse duration, and repetition rate of the pulse laser used in this study were 800 nm, $\sim $ 120 fs, and 1 kHz, respectively. 
The pump pulse was irradiated on the film at an angle of 10 degrees measured from the film normal. 
The polar MOKE of the probe laser pulse was detected to measure the magnetization component normal to the film plane. 
To detect the pump pulse induced change in the polar MOKE signal, the pump pulse was modulated with a frequency of 360 Hz by using a mechanical chopper. The pump induced change in the MOKE signal was detected by a lock-in amplifier. 
The pump fluence $F_{\rm p}$ used in this study was fixed to 8.3 J/m$^2$. 
An in-plane external magnetic field up to 2 T was applied.

\section{Results and discussions}

\subsection{Optical helicity induced magnetization dynamics and its magnetic field dependence}

Figure 1(b) shows typical femtosecond laser pulse induced magnetization dynamics where left circularly polarized (LCP) laser and right circularly polarized (RCP) laser were irradiated and a 2 T in-plane magnetic field was applied. 
A change in the initial phase of magnetization precession was observed when light of different optical helicity was irradiated on the film. 
A slight difference was observed in precessional amplitude owing to sample misalignment. 
This is because thermal demagnetization can induce magnetization precession if the magnetic field is slightly tilted toward the out-of-plane direction. 
The signal at approximately $t=0$ could be due to the fact that MOKE senses the orbital magnetic moment excited by the optical helicity pulse, the so-called specular inverse Faraday effect and specular inverse optical Kerr effect \cite{Wilks2003, Kruglyak2005, Longa2007, Choi2019}. 
Figure 2(a) shows the magnetic field dependence of the optical helicity-induced magnetization precession in the MgO/FeCo(2)/Pt(3) thin film.
The magnetization precession dynamics after the initial peak were fitted using a sinusoidal decay function as $m_0 \exp \left( -t/\tau \right) \sin (2\pi f t + \varphi _0) $, where $m_0, f, \tau , \varphi _0$ are amplitude, frequency, life-time, and initial phase of magnetization precession, respectively.  
The precession frequency increases with an increase in the magnetic field, while the amplitude does not significantly depend on the magnetic field. 
The precession frequency $f$ is evaluated by the fitting plotted as a function of the magnetic field $\mu _0 H$ in Fig. 2(b).
The $f$ vs $\mu _0 H$ curve is fitted by the Kittel formula as, $f = \gamma  \sqrt{\mu_0 H(\mu_0 H+\mu_0 M_{\rm eff})}/ (2\pi )$, where $\gamma $ and $\mu _0 M_{\rm eff}$ are gyromagnetic ratio and an effective demagnetization field.
Figure 2(c) shows the inverse lifetime $1/\tau $ evaluated by the fitting plotted as a function of $\mu _0 H$. The $1/\tau $ vs $\mu _0 H$ plot is well explained by the theoretical relation $1/\tau = \alpha \gamma (\mu _0 H+\mu _0 M_{\rm eff} /2)$, where the Gilbert damping parameter $\alpha $ is 0.019.
The large value of $\alpha $ compared with $\alpha $ for bulk FeCo ($\sim $ 0.002 \cite{Schoen2016}) is due to the spin-pumping effect at the FeCo/Pt interface \cite{Tserkovnyak2002}.
Figure 2(d) shows the averaged precession amplitude ($(|m_0({\rm LCP})|+ |m_0({\rm RCP})|)/2$) plotted as a function of precession frequency.
The precession frequency does not have a significant influence on the precession amplitude, indicating non-thermal ultrafast pulse-like excitation induced by the optical spin torque. 

\subsection{Thickness dependences of the optical helicity induced magnetization precession}

To understand the excitation mechanism of the optical helicity induced magnetization precession in FeCo/Pt bilayer, nonmagnetic Pt and ferromagnetic FeCo thickness dependence were measured as shown in Fig. 3. 
To exclude the effect of the sample misalignment as mentioned above, we measured the signal for both + 2 T and – 2 T, and the average was obtained. In addition, the MOKE signal was normalized by the static MOKE voltage to detect the change in normalized magnetization.
Figures 3(a) and 3(b) show the Pt thickness $d_{\rm Pt}$ and FeCo thickness $d_{\rm FeCo}$ dependences of the optical helicity induced magnetization precession. 
It was found that the precession amplitude initially increased with increasing $d_{\rm Pt}$ from 2 to 5 nm and then slightly decreased at $d_{\rm Pt}$ = 10 nm. 
The decrease in amplitude with increasing $d_{\rm FeCo}$ [Fig. 3(b)] can be qualitatively explained by the fact that the optical spin torque was induced by interface effects.

\subsection{Evaluation of optical spin torque vectors}

To evaluate the optical spin vector quantitatively for different thicknesses, the phase of the magnetization precession was analyzed as shown in Fig. 4.
Figure 4(a) shows the typical time-resolved MOKE signal when a circularly polarized laser pulse was irradiated on the sample. The data were obtained by taking the differences between signals measured with LCP and RCP lights. All data for the phase analysis are provided in Supplemental Material \cite{Sup}. The large peak due to specular inverse Faraday effect, as mentioned above, is used to define $t$ = 0, the time when pump pulse arrives.
Figure 4(b) shows the subsequent magnetization precession.
The signal is decomposed into the cosine and sine components by using the following fitting function:
\begin{align}
f(t) = \left(a_1 \cos (2\pi f t)+a_2 \sin (2\pi f t) \right) \exp (-t/\tau ) , \label{eq:fit}
\end{align}
where $a_1$ and $a_2$ are the cosine and sine component amplitudes, respectively.
They are shown as red and blue solid curves in Fig. 4(b), respectively. 
If the magnetization precession is assumed to be induced by the out-of-plane spin-transfer torque (${\bf m}\times ({\bf m} \times {\bf s})$), the magnetization is simultaneously tilted away from the film plane by the femtosecond laser pulse and the precession starts from the magnetization tilted away from the film plane, which leads to a cosine signal in the out-of-plane component of magnetization. 
On the other hand, if magnetization precession is excited by the field-like torque (${\bf m}\times {\bf s}$), magnetization is tilted toward in-plane and precession starts subsequently, leading to a sine signal.
Note that \color{black}the easy axis of the magnetization is in the film plane and the magnetization is parallel to the magnetic field direction in equilibrium. 
\color{black}
A small in-plane component of the light wavevector is parallel to the magnetization and does not induce torque on the magnetization.
The magnetization dynamics excited by the optical spin torque can be described by the Landau-Lifshitz-Gibert equation as follows,
\begin{align}
\frac{d{\bf m}}{dt} = -\gamma \mu _0  {\bf m} \times {\bf H}_{\rm eff} + \alpha {\bf m} \times \frac{d{\bf m}}{dt} - \tau _{\parallel}(t) {\bf m}\times {\bf z} - \tau _{\perp }(t) {\bf m}\times ({\bf m} \times {\bf z}) , \label{eq:LLG}
\end{align} 
where, the first and second terms are precession torque and damping torque.
$\tau _{\perp }$ and $\tau _{\parallel }$ are the out-of-plane and in-plane optical spin torques, respectively. 
${\bf H}_{\rm eff}$ is a effective magnetic field due to external magnetic field and anisotropy field, which is given by,
\begin{align}
{\bf H}_{\rm eff} = {\bf H} - M_{\rm eff} ({\bf m}\cdot {\bf z}) {\bf z}.
\end{align}
By solving the linearized Landau-Lifshitz-Gilbert equation, the following equation is obtained for the out-of-plane component of the magnetization:
\begin{align}
m_{\rm z}(t) &= \left( \tau _{\perp } \cos (2\pi ft ) + \sqrt{\frac{H}{H+M_{\rm eff}}} \tau _{\parallel } \sin (2\pi f t) \right) \Delta t \notag \\
&\hspace{4cm} \times \exp (-t /\tau ) , \label{eq:mz}
\end{align} 
where $m_{\rm z}$ is the out-of-plane component of the normalized magnetization.
Here, the phase delay due to the damping term was neglected since $\alpha $ is small enough ($\tan ^{-1} (\alpha )$ = 0.6 $ {\rm deg.} $ when $\alpha $ = 0.01 is used).
$\Delta t$ is the pulse duration, and it is considerably shorter than the period of the magnetization precession, i.e, $2\pi f \Delta t \ll 1 $.  
Therefore, the time-integrated out-of-plane torque $\tau _{\perp }\Delta t$ and in-plane torque $\tau _{\parallel }\Delta t$ can be evaluated by using Eqs. (\ref{eq:fit}) and (\ref{eq:mz}).
It should be noted that the effect of $\tau _{\parallel }$ on $m_{\rm z}$ is increased with increasing $H$ because the elliptical shape of magnetization precession is suppressed by the bias magnetic field.
Figure 4(c) and 4(d) show the two-dimensional plots of the optical spin torques, $\tau _{\perp }\Delta t$ and $\tau _{\parallel }\Delta t$.
It was found that the torque vectors depend on both $d_{\rm Pt}$ and $d_{\rm FeCo}$.
The $d_{\rm Pt}$ and $d_{\rm FeCo}$ dependences of the out-of-plane torque and in-plane torque are summarized in Fig. 4(e) and 4(f).

\subsection{Discussions of the thickness dependence of the optical spin torques}

To discuss the optical spin torque with different $d_{\rm Pt}$ and $d_{\rm FeCo}$, light absorption and Poynting vectors are calculated with the refractive index values taken from the literature \cite{Werner2009, Stephens1952, Malitson1965, Green2008}. 
The details of the calculation are provided in the Supplemental Material \cite{Sup} (see Supplemental Material, Sec. I and I\hspace{-.01em}I).
The out-of-plane and in-plane torques are discussed based on the optically generated spin and spin-transport in bilayer thin films, analogous to electrical generation of spin-orbit torque \cite{Haney2013, Amin2016, Miron2010, Kim2012, Kurebayashi2014, Hayashi2014, Avci2014, Pai2014, Emori2016, Manchon2009, Abiague2009}.
The $d_{\rm Pt}$ and $d_{\rm FeCo}$ dependences of out-of-plane torque can be understood by the fact that spin generated by the optical orientation in Pt layer induces out-of-plane spin-transfer torque. 
The $d_{\rm Pt}$ dependence of out-of-plane torque shown in Fig. 4(e) is similar to  light absorption in Pt layer [dashed curve in Fig. 4(e)], indicating that spin is generated by light absorption in Pt layer. 
In addition, the out-of-plane torque increased with decreasing $d_{\rm FeCo}$, as indicated in Fig. 4(f). 
This is consistent with the nature of the interface effect of the spin-transfer torque. 
The spin angular momentum conversion efficiency in Pt layer calculated by using the light absorption in Pt layer was found to be \color{black}approximately 2\% (see Supplemental Material, sec. V \cite{Sup}).
This value is roughly consistent with that reported previously\cite{Choi2020}.
\color{black}

The in-plane torque can be potentially induced by the following three mechanisms: spin-rotation at the interface, inverse Faraday effect in the ferromagnetic layer, and spin-generation via interface spin-orbit coupling.
The spin is slightly rotated when it travels across the nonmagnet/ferromagnet interface, which can be described by the imaginary part of the spin-mixing conductance at the interface \cite{Haney2013, Amin2016}. 
However, if the in-plane torque is assumed to be due to the imaginary part of the spin-mixing conductance, the $d_{\rm Pt}$ dependence of in-plane torque should exhibit the same behavior as the out-of-plane torque. 
In addition, the imaginary part of the spin-mixing conductance is smaller than the real part by a factor greater than ten \cite{QZhang2011}. 
These facts rule out the possibility of spin-rotation at the interface.
The second possible consideration is that the in-plane field-like torque is induced by the magnetic field generated by the inverse Faraday effect.
In the inverse Faraday effect, the magnetic field generated by the circularly polarized light is proportional to the square of the electric field amplitude $E\times E^{\star }$. 
\color{black}
However, the significant increase of the in-plane torque with decreasing FeCo thickness cannot be explained by the inverse Faraday effect inside the FeCo layer (see Supplemental Material, sec I\hspace{-.1em}V). 
\color{black}
Hence, the above fact cannot explain the $d_{\rm FeCo}$ dependence of in-plane torque[Fig. 4(f)] and rules out the second scenario.

The remaining possible mechanism is that the in-plane torque is induced by the spin generated due to the interface spin-orbit coupling, because the in-plane field-like torque is inversely proportional to $d_{\rm FeCo}$ [Solid line in Fig. 4(f)].
If the stacking structure exhibits inversion symmetry breaking, spin is generated due to Rashba spin-orbit coupling. 
Optically induced spin via Rashba spin-orbit coupling has been theoretically discussed in previous studies \cite{Edelstein1997, Taguchi2012, Qaiu2016, Li2017, Mochizuki2018}.
Here, we consider the Rashba spin-orbit coupling with inversion symmetry breaking to be normal to the film plane,
\begin{align}
\mathcal{H}_{\rm R} =\frac{\alpha _{\rm R}}{\hbar } \left( {\boldsymbol \sigma } \times {\bf p} \right) \cdot {\bf z},
\end{align}
where $\alpha _{\rm R}$, ${\boldsymbol \sigma }$, ${\bf p}$, and ${\bf z}$ are the Rashba parameter, Pauli spin matrix, electron momentum, and unit vector normal to film plane, respectively.
Edelstein derived the optically induced spin due to Rashba spin-orbit coupling as \cite{Edelstein1997},
\begin{align}
{\bf s}_{\rm R} = i K {\bf z} \left( {\bf z} \cdot {\bf E} \times {\bf E}^{\star } \right) , \label{eq:sR}
\end{align}
where we used $K$ as a coefficient of the induced spin ${\bf s}_{\rm R}$.
Note that the spin is generated parallel to the film normal, and its sign does not depend on the polarity of the symmetry.
The exchange coupling between the magnetization and the spin generated via Rashba spin-orbit coupling induces the in-plane torque ${\bf m}\times {\bf s}_{\rm R}$ \cite{Manchon2009, Abiague2009}, which is analogous to the electrical manipulation of magnetization via the Rashba-Edelstein effect.

To validate the optical Rashba--Edelstein effect, the optical helicity-induced magnetization precession with a changing stacking structure symmetry (Pt/Co/Pt, Pt/Co/Ta, and Co/Pt) was measured \cite{Sup} (see Supplemental Material, sec. V\hspace{-.1em}I\hspace{-.1em}I\hspace{-.1em}I).
The in-plane torque was enhanced when Co was sandwiched by SiO$_2$ and Pt. This confirms that the in-plane torque is induced by the interface Rashba effect due to the asymmetric heterostructure. 
In addition, the electrical interfacial field-like spin-orbit torque due to the Rashba--Edelstein effect has been reported in similar stacking structures (oxide/(Co, CoFe, CoFeB)/Pt ) \cite{Hayashi2014, Pai2014, Du2020}. This supports the presence of Rashba spin-orbit coupling in the heterostructures used in this study.

\color{black}
Finally, the in-plane torque $\tau _{\parallel } \Delta t$ obtained from the experiments are discussed quantitatively in terms of optical Rashba--Edelstein effect.
The in-plane torque in the unit of the angular momentum per unit area $M_{\rm s}d_{\rm FeCo} \tau _{\parallel }\Delta t /\gamma $, which was obtained by the slope of $\tau _{\parallel }\Delta t $ vs $1/d_{\rm FeCo}$ in Fig. 4(f), was evaluated to be $\sim $ 4 $\times $ 10$^{-18}$ $[{\rm J \cdot s \cdot m}^{-2}]$.
This value agrees with the calculated value $\sim 2$ $\times $ 10$^{-18}$ $[{\rm J \cdot s \cdot m}^{-2}]$\cite{Sup} (see Supplemental Material, sec. V\hspace{-.1em}I\hspace{-.1em}I), in which the Rashba parameter and the exchange coupling constant are taken from literature\cite{Miron2010, Ast2007, Barreteau2004}.  

\color{black}

\section{Conclusion}

This study investigated the optical spin torque vector in FeCo/Pt heterostructure with different thicknesses. 
$d_{\rm Pt}$ and $d_{\rm FeCo}$ dependences of the out-of-plane and in-plane optical spin torque were evaluated by analyzing the magnetization precession phase. It was shown that the in-plane field-like optical spin torque significantly increased with decreasing $d_{\rm FeCo}$, indicating the interfacial nature of the torque. 
The in-plane field-like torque was induced by the spin generated due to the interface spin-orbit coupling as a result of the optical Rashba--Edelstein effect. 
The optical generation of spin via optical Rashba-Edelstein effect represents a new method of manipulating thin film nanomagnets. 
However, the field-like optical spin torque induced by the interfacial spin is small compared with the spin-transfer torque induced by the optical orientation of Pt. 
The materials that are known to show Rashba interface, such as Ag/Bi interface \cite{Ast2007, Sanchez2013, Nakayama2016}, may prove valuable in increasing the efficiency of the interface optical spin generation.

\begin{acknowledgments}
This work was partially supported by KAKENHI (No. 26103004), Advanced Technology Institute Research Grants, the Center for Spintronics Research Network (CSRN), and the JSPS Core-to-Core Program.
\end{acknowledgments}

\begin{figure}[t]
\begin{center}
\includegraphics[width=10cm,keepaspectratio,clip]{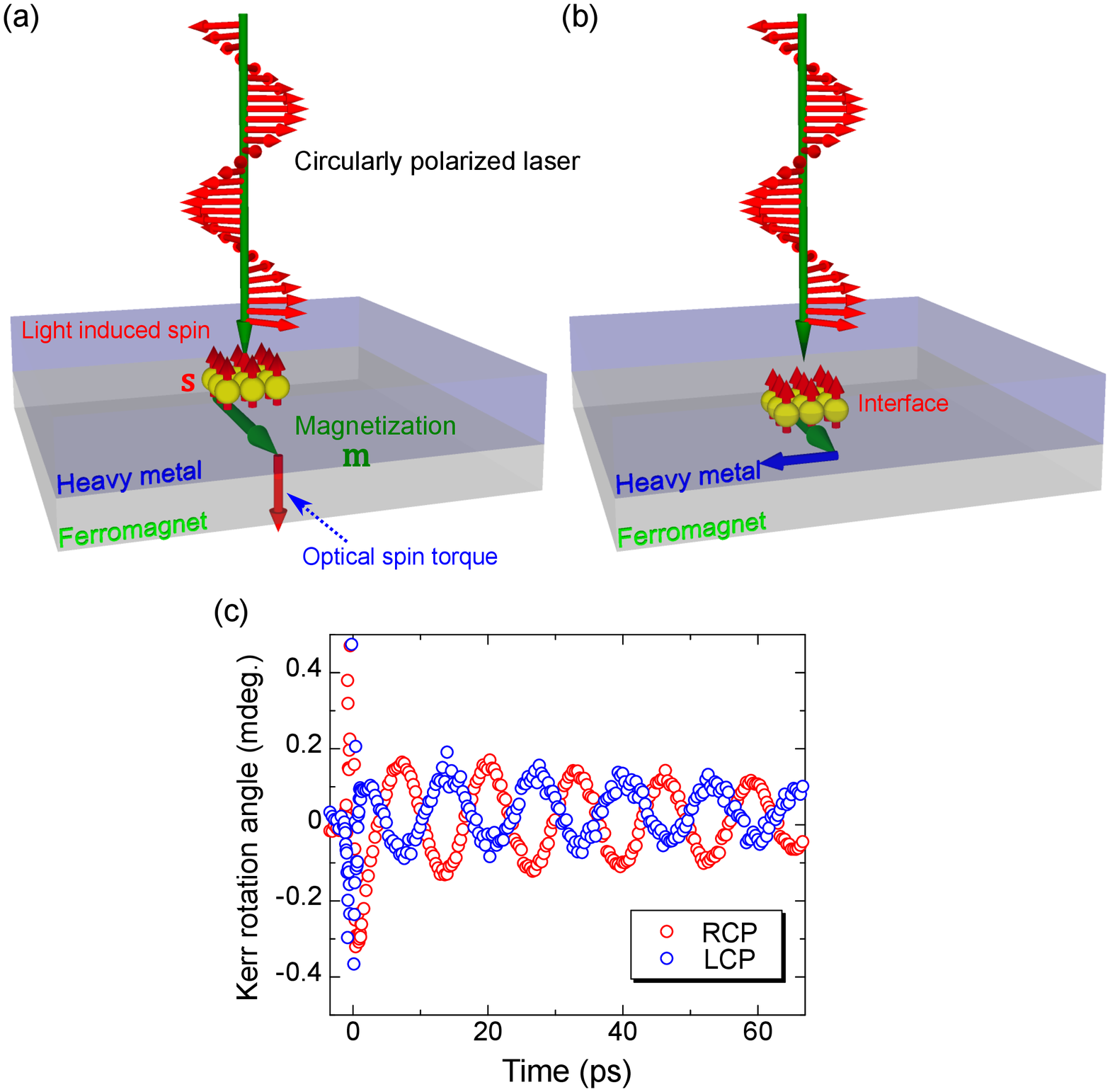}
\end{center}
\caption{\color{black}(a) Schematic illustration of optical spin torque induced by the optical orientation in the heavy metal layer. (b) Field-like optical spin torque induced by the circularly polarized laser due to interface spin-orbit coupling.\color{black} (c) Typical magnetization precessional dynamics excited by circularly polarized laser pulse with different optical helicity for 2-nm-thick FeCo/3-nm-thick Pt thin film. }
\label{f1}
\end{figure}

\begin{figure}[t]
\begin{center}
\includegraphics[width=10cm,keepaspectratio,clip]{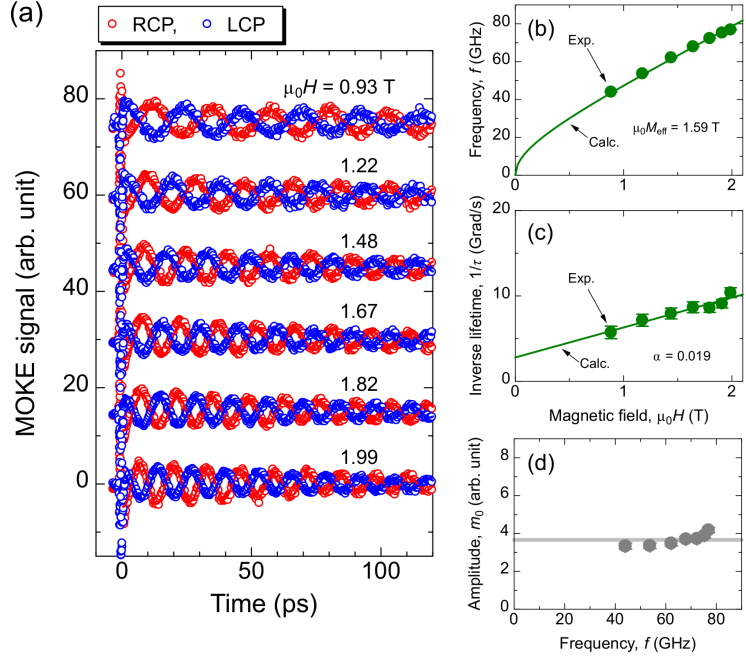}
\end{center}
\caption{(a) Circularly polarized laser pulse induced magnetization precession in 2-nm-thick FeCo/3-nm-thick Pt bilayer with different magnetic fields $\mu _0 H$. (b) Precession frequency and (c) inverse lifetime obtained by fitting plotted as a function of magnetic field. Solid curves are the results calculated using theoretical formula. (d) Precession amplitude plotted as a function of precession frequency. Horizontal line is a guide to eye.}
\label{f2}
\end{figure}

\begin{figure}[t]
\begin{center}
\includegraphics[width=10cm,keepaspectratio,clip]{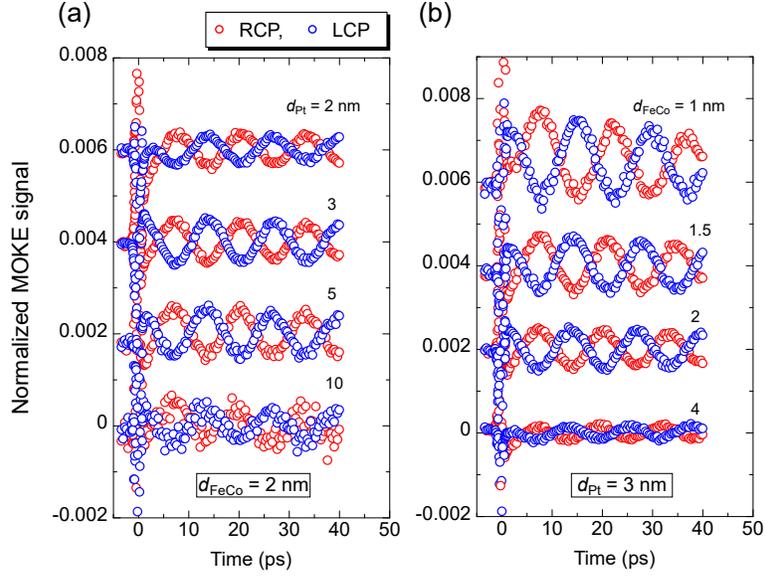}
\end{center}
 \caption{(a) Pt and (b) FeCo thickness dependence of optical helicity induced magnetization precession for FeCo/Pt bilayer. The sample ($d_{\rm Pt}, d_{\rm FeCo}$) = (2 nm, 3 nm) was measured twice.}
\label{f3}
\end{figure}

\begin{figure}[t]
\begin{center}
\includegraphics[width=12cm,keepaspectratio,clip]{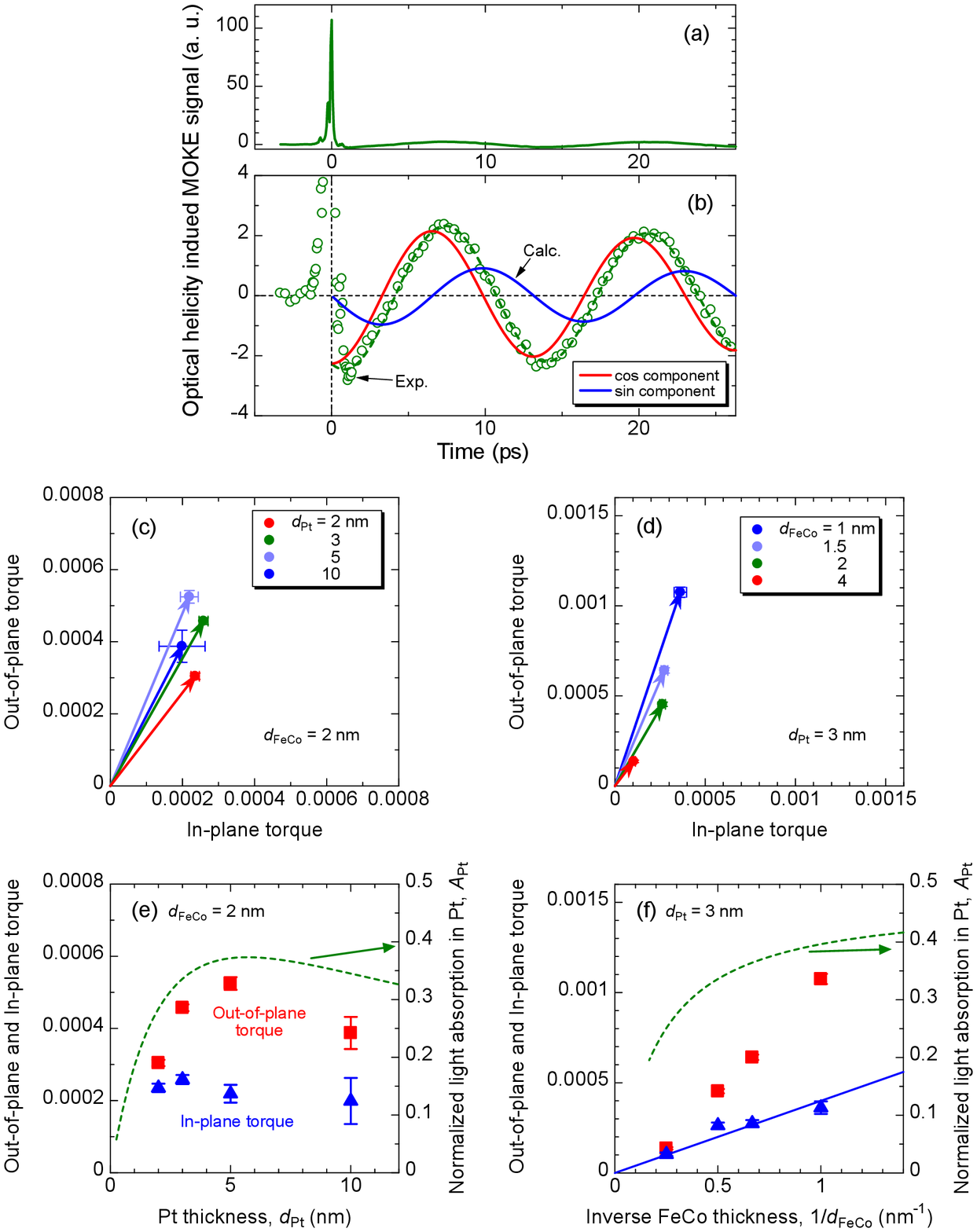}
\end{center}
\caption{(a) MOKE signal as a function of pump-probe delay time when circularly polarized laser pulse was irradiated on FeCo/Pt thin films. The optical helicity induced signal was observed at zero-delay time. (b) Subsequent optical helicity induced magnetization precession. Dashed curves are fitted results and decomposed into cosine and sine components, shown as red and blue solid curves, respectively. (c) Pt and (d) FeCo thickness dependence of two-dimensional optical spin torque in FeCo/Pt thin films extracted from the phase analysis. (e) Pt and (f) FeCo thickness dependence of the out-of-plane [square symbols] and in-plane torque [triangle symbols]. Dashed curves are calculated normalized light absorption in Pt layer. Solid line in (f) represents in-plane torque proportional to 1/$d_{\rm FeCo}$. }
\label{f5}
\end{figure}


\begin{thebibliography}{99}

%----------------Spin-Hall effect and Rashba torque---------------
\bibitem{Manchon2015}
A. Manchon, H. C. Koo, J. Nitta, S. M. Frolov, R. A. Duine,
Nat. Mater. {\bf 14}, 871 (2015).
\bibitem{Miron_Nature}
I. M. Miron, K. Garello, G. Gaudin, P. -J. Zermatten, M. V. Costache, S. Auffret, S. Bandiera, A. Schuhl, P. Gambardella,
Nature, {\bf 476}, 189 (2011).
\bibitem{Liu_Science}
L. Liu, C. -F. Pai, Y. Li, H. W. Tseng, D. C. Ralph, R. A. Buhrman,
Science, {\bf 336}, 555 (2012).

%-----------Theory of spin-orbit torque--------
\bibitem{Haney2013}
P. M. Haney, H. -W. Lee, K. -J. Lee, A. Manchon, M. D. Stiles,
Phys. Rev. B {\bf 87}, 174411 (2013).
\bibitem{Amin2016}
V. P. Amin, M. D. Stiles,
Phys. Rev. B {\bf 94}, 104420 (2016).


%----------------All-optical helicity dependent switching----------------
\bibitem{Lambert2014}
C. -H. Lambert, S. Mangin, B. S. D. Ch. S. Varaprasad, Y. K. Takahashi, M. Hehn, M. Cinchetti, G. Malinowski, K. Hono, Y. Fainman, M. Aeschlimann, E. E. Fullerton, 
Science, {\bf 345}, 1337 (2014).
\bibitem{Hadri2016}
M. S. El Hadri, M. Hehn, P. Pirro, C. -H. Lambert, G. Malinowski, E. E. Fullerton, S. Mangin,
Phys. Rev. B {\bf 94}, 064419 (2016).
\bibitem{Takahashi2016}
Y. K. Takahashi, R. Medapalli, S. Kasai, J. Wang, K. Ishioka, S. H. Wee, O. Hellwig, K. Hono, E. E. Fullerton,
Phys. Rev. Appl. {\bf 6}, 054004 (2016).
\bibitem{John2017}
R. John, M. Berritta, D. Hinzke, C. M\"{u}ller, T. Santos, H. Ulrichs, P. Nieves, J. Walowski, R. Mondal, O. Chubykalo-Fesenko, J. McCord, P. M. Oppeneer, U. Nowak, M. M\"{u}nzenberg.
Sci. Rep. 7, 4114 (2017).
\bibitem{Kichin2019}
G. Kichin, M. Hehn, J. Gorchon, G. Malinowski, J. Hohlfeld, S. Mangin,
Phys. Rev. Appl. {\bf 12}, 024019 (2019).

%---------------theoretical calculation--------------
\bibitem{Battiato2014}
M. Battiato, G. Barbalinardo, P. M. Oppeneer,
Phys. Rev. B {\bf 89}, 014413 (2014).
\bibitem{Berritta2016}
M. Berritta, R. Mondal, K. Carva, P. M. Oppeneer, 
Phys. Rev. Lett. {\bf 117}, 137203 (2016).
\bibitem{Freimuth2016}
F. Freimuth, S. Bl\"{u}gel, Y. Mokrousov,
Phys. Rev. B {\bf 94}, 144432 (2016).

%---------------opto-spintronic device--------------
\bibitem{Kimel2019}
A. V. Kimel, M. Li,
Nat. Rev. Mater. {\bf 4} 189 (2019).
\bibitem{Becker2020}
H. Becker, C. J. Kr\"{u}ckel, D. V. Thourhout, M. J. R. Heck,
IEEE J. Sel. Top. Quant. {\bf 26}, 8300408 (2020).

%---------------MCD---------------------
\bibitem{Ellis2016}
M. O. A. Ellis, E. E. Fullerton, R. W. Chantrell,
Sci. Rep. {\bf 6}, 30522 (2016)
\bibitem{Gorchon2016}
J. Gorchon, Y. Yang, J. Bokor,
Phys. Rev. B {\bf 94}, 020409(R) (2016).

%--------------Helicity induced magnetization precession--------
\bibitem{Choi2017}
G. -M. Choi, A. Schleife, D. G. Cahill,
Nat. Commun. {\bf 8}, 15085 (2017).
\bibitem{Choi2020}
G. -M. Choi, J. H. Oh, D. -K. Lee, S. -W. Lee, K. W. Kim, M. Lim, B. -C. Min, K. -J. Lee, H. W. Lee,
Nat. Commun, {\bf 11}, 1482 (2020).
\bibitem{Choi2019}
G. -M. Choi, H. G. Park, B. -C. Min,
J. Magn. Magn. Mater. {\bf 474}, 132 (2019).

%--------------Terahertz emission-------------------
\bibitem{Huisman2016}
T. J. Huisman, R. V. Mikhaylovskiy, J. D. Costa, F. Freimuth, E. Paz, J. Ventura, P. P. Freitas, S. Bl\"{u}gel, Y. Mokrousov, Th. Rasing, A. V. Kimel,
Nat. Nanotech. {\bf 11}, 455 (2016).
\bibitem{Jungfleish2018}
M. B. Jungfleisch, Q. Zhang, W. Zhang, J. E. Pearson, R. D. Schaller, H. Wen, A. Hoffmann,
Phys. Rev. Lett. {\bf 120}, 207207 (2018).

%--------------Insulator or Semiconductor----------------
\bibitem{Lampel1967}
G. Lampel,
Phys. Rev. Lett. {\bf 20}, 491, (1967).
\bibitem{Awschalom1986}
D. D. Awschalom, J. Warnock, S. von Moln\'{a}r,
Phys. Rev. Lett. {\bf 58}, 812 (1987).
\bibitem{Oiwa2002}
A. Oiwa, Y. Mitsumori, R. Moriya, T. Slupinski, H. Munekata,
Phys. Rev. Lett. {\bf 88}, 137202 (2002).
\bibitem{Kimel2005}
A. V. Kimel, A. Kirilyuk, P. A. Usachev, R. V. Pisarev, A. M. Balbashov, Th. Rasing,
Nature {\bf 435}, 655 (2005).
\bibitem{Satoh2010}
T. Satoh, S. -J. Cho, R. Iida, T. Shimura, K. Kuroda, H. Ueda, Y. Ueda, B. A. Ivanov, F. Nori, M. Fiebig,
Phys. Rev. Lett. {\bf 105}, 077402 (2010).
\bibitem{Nemec2012}
P. N\u{e}mec, E. Rozkotov\'{a}, N. Tesa\u{r}ov\'{a}, F. Troj\'{a}nek, E. De Ranieri, K. Olejn\'{i}k, J. Zemen, V. Nov\'{a}k, M. Cukr, P. Mal\'{y}, and T. Jungwirth,
Nat. Phys. {\bf 8}, 411 (2012).
\color{black}
\bibitem{Tesarova2013}
N. Tesa\u{r}ov\'{a}, P. N\u{e}mec, E. Rozkotov\'{a}, J. Zemen, T. Janda, D. Butkovi\u{c}ov\'{a}, F. Troj\'{a}nek, K. Olejnik, V. Nov\'{a}k, P. Mal\'{y}, T. Jungwirth,
Nat. Photo. {\bf 7}, 492 (2013).
\color{black}
\bibitem{Ramsay2015}
A. J. Ramsay, P. E. Roy, J. A. Haigh, R. M. Otxoa, A. C. Irvine, T. Janda, R. P. Campion, B. L. Gallagher, J. Wunderlich,
Phys. Rev. Lett. {\bf 114}, 067202 (2015).

%----------Edelstein1997----------
\bibitem{Edelstein1997}
V. M. Edelstein,
Phys. Rev. Lett. {\bf 80}, 5766 (1998).

%----------specular inverse effect----------
\bibitem{Wilks2003}
R. Wilks, N. D. Hughes, R. J. Hicken,
J. Phys.: Condens. Matter {\bf 15}, 5129 (2003).
\bibitem{Kruglyak2005}
V. V. Kruglyak, R. J. Hicken, M. Ali, B. J. Hickey, A. T. G. Pym, B. K. Tanner,
Phys. Rev. B {\bf 71}, 233104 (2005).
\bibitem{Longa2007}
F. Dalla Longa, J. T. Kohlhepp, W. J. M. de Jonge, B. Koopmans,
Phys. Rev. B {\bf 75}, 224431 (2007).


%-----------measurement system--------
\bibitem{Iihama2014}
S. Iihama, S. Mizukami, H. Naganuma, M. Oogane, Y. Ando, T. Miyazaki,
Phys. Rev. B {\bf 89}, 174416 (2014).
\bibitem{Mizukami2016}
S. Mizukami, S. Iihama, Y. Sasaki, A. Sugihara, R. Ranjibar, K. Z. Suzuki,
J. Appl. Phys. {\bf 120}, 142102 (2016).

%-----------Damping parameter----------
\bibitem{Schoen2016}
M. A. W. Schoen, D. Thonig, M. L. Schneider, T. J. Silva, H. T. Nembach, O. Eriksson, O. Karis, J. M. Shaw,
Nat. Phys. {\bf 12}, 839 (2016).
\bibitem{Tserkovnyak2002}
Y. Tserkovnyak, A. Brataas, G. E. W. Bauer,
Phys. Rev. Lett. {\bf 88}, 117601 (2002).


%-----refractive index-----
\bibitem{Werner2009}
W. S. M. Werner,
J. Phys. Chem. Ref. {\bf 38}, 1013 (2009).
\bibitem{Stephens1952}
R. E. Stephens, I. H. Malitson,
J. Res. Natl. Bur. Stand. {\bf 49}, 249 (1952).
\bibitem{Malitson1965}
I. H. Malitson,
J. Opt. Soc. Am. {\bf 55}, 1205 (1965).
\bibitem{Green2008}
M. A. Green,
Sol. Ener. Mater. Sol. Cells {\bf 92}, 1305 (2008).


%-----------Supplemental material--------
\bibitem{Sup}
See Supplemental Material for light absorption calculation, calculation of inverse Faraday effect, evaluation of spin-angular momentum conversion efficiency in Pt, all data of phase analysis, measurement results for Pt/Co/Pt, Pt/Co/Ta, Co/Pt samples, \color{black}and calculation results of in-plane torque due to Rashba spin-orbit coupling\color{black}.



%-----------Study on electrical manipulation of spin-orbit torque----------
\bibitem{Miron2010}
I. M. Miron, G. Gaudin, S. Auffret, B. Rodmacq, A. Schuhl, S. Pizzini, J. Vogel, P. Gambardella,
Nat. Mater. {\bf 9}, 230 (2010).
\bibitem{Kim2012}
J. Kim, J. Sinha, M. Hayashi, M. Yamanouchi, S. Fukami, T. Suzuki, S. Mitani, H. Ohno,
Nat. Mater. {\bf 12}, 240 (2012).
\bibitem{Kurebayashi2014}
H. Kurebayashi, J. Sinova, D. Fang, A. C. Irvine, T. D. Skinner, J. Wunderlich, V. Nov\'{a}k, R. P. Campion, B. L. Gallagher, E. K. Vehstedt, L. P. Z\^{a}rbo, K. V\'{y}born\'{y}, A. Ferguson, T. Jungwirth,
Nat. Nanotech. {\bf 9}, 211 (2014).
\bibitem{Hayashi2014}
M. Hayashi, J. Kim, M. Yamanouchi, H. Ohno,
Phys. Rev. B {\bf 89}, 144425 (2014).
\bibitem{Avci2014}
C. O. Avci, K. Garello, M. Gabureac, A. Ghosh, A. Fuhrer, S. F. Alvarado, P. Gambardella,
Phys. Rev. B {\bf 90}, 224427 (2014).
\bibitem{Pai2014}
C. -F. Pai, Y. Ou, L. H. Vilela-Le\~{a}o, D. C. Ralph, R. A. Buhrman,
Phys. Rev. B {\bf 92}, 064426 (2015).
\bibitem{Emori2016}
S. Emori, T. Nan, A. M. Belkessam, X. Wang, A. D. Matyushov, C. J. Babroski, Y. Gao, H. Lin, N. X. Sun,
Phys. Rev. B {\bf 93}, 180402(R) (2016).
\bibitem{Manchon2009}
A. Manchon, S. Zhang,
Phys. Rev. B {\bf 79}, 094422 (2009).
\bibitem{Abiague2009}
A. Matos-Abiague, R. L. Rodr\'{i}guez-Su\'{a}rez,
Phys. Rev. B {\bf 80}, 094424 (2009).

%--------Mixing conductance---------
\bibitem{QZhang2011}
Q. Zhang, S. Hikino, S. Yunoki,
Appl. Phys. Lett. {\bf 99}, 172105 (2011).


%--------Theory of optical spin torque using Rashba effect--------

\bibitem{Taguchi2012}
K. Taguchi, G. Tatara,
J. Phys. Conf. {\bf 400}, 042055 (2012).
\bibitem{Qaiu2016}
A. Qaiumzadeh, M. Titov,
Phys. Rev. B {\bf 94}, 014425 (2016).
\bibitem{Li2017}
J. Li, P. M. Haney,
Phys. Rev. B {\bf 96}, 054447 (2017).
\bibitem{Mochizuki2018}
M. Mochizuki, K. Ihara, J. Ohe, A. Takeuchi,
Appl. Phys. Lett. {\bf 112}, 122401 (2018).

%--------Parameters-------
\color{black}
\bibitem{Ast2007}
G. R. Ast, J. Henk, A. Ernst, L. Moreschni, M. C. Falub, D. Pacil\'{e}, P. Bruno, K. Kern, M. Grioni,
Phys. Rev. Lett. {\bf 98}, 186807 (2007).
\bibitem{Barreteau2004}
C. Barreteau, M. -C. Desjonqueres, A. M. Ole\'{s}, D. Spanjaard,
Phys. Rev. B {\bf 69}, 064432 (2004).
\color{black}

%---------Rashba effect in Co(Fe)-Pt / insulator structure--------
\bibitem{Du2020}
Y. Du, H. Gamou, S. Takahashi, S. Karube, M. Kohda, J. Nitta,
Phys. Rev. Appl. {\bf 13}, 054014 (2020).

%---------Bi / Ag Rashba interface----------
\bibitem{Sanchez2013}
J. C. R. S\'{a}nchez, L. Vila, G. Desfonds, S. Gambarelli, J. P. Attan\'{e}, J. M. De Teresa, C. Mag\'{e}n, A. Fert,
Nat. Commun. {\bf 4}, 2944 (2013).
\bibitem{Nakayama2016}
H. Nakayama, Y. Kanno, H. An, T. Tashiro, S. Haku, A. Nomura, K. Ando,
Phys. Rev. Lett. {\bf 117}, 116602 (2016).


\end{thebibliography}
\end{document}

\title{
Supplemental Material: \\ Interface-induced field-like optical spin torque in a ferromagnet/heavy metal heterostructure}

\author{Satoshi Iihama}
\affiliation{Frontier Research Institute for Interdisciplinary Sciences (FRIS), Tohoku University, Sendai 980-8578, Japan}
\affiliation{Center for Spintronics Research Network (CSRN), Tohoku University, Sendai 980-8577, Japan}

\author{Kazuaki Ishibashi}
\affiliation{Department of Applied Physics, Graduate School of Engineering, Tohoku University, 6-6-05, Aoba-yama, Sendai 980-8579, Japan}
\affiliation{WPI Advanced Institute for Materials Research (AIMR), Tohoku University, 2-1-1, Katahira, Sendai 980-8577, Japan}

\author{Shigemi Mizukami}
\affiliation{Center for Spintronics Research Network (CSRN), Tohoku University, Sendai 980-8577, Japan} 
\affiliation{WPI Advanced Institute for Materials Research (AIMR), Tohoku University, 2-1-1, Katahira, Sendai 980-8577, Japan}
\affiliation{Center for Science and Innovation in Spintronics (CSIS), Core Research Cluster (CRC), Tohoku University, Sendai 980-8577, Japan}

\date{\today}

\maketitle

\section{Refractive index}

The reflectance, light absorption, and Poynting vector in multilayer thin films are calculated via a transfer matrix method\cite{Byrnes2016}. The reflectance was observed to increase with both FeCo and Pt thickness, as shown in Figs. S1(a) and S1(b). The experimentally obtained values are roughly consistent with the values calculated using the transfer matrix and refractive indices summarized in Table SI. The refractive index for Fe--Co alloy is obtained by taking the average of the refractive indices of Fe and Co. 

\begin{table*}[ht]
\caption{Refractive index (800 nm) used for transfer matrix calculation}
\begin{tabular}{cccc}\hline \hline
\hspace{0.5cm} Material \hspace{0.5cm} & \hspace{0.5cm} thickness (nm) \hspace{0.5cm} & \hspace{0.5cm} Refractive index\hspace{0.5cm} & \hspace{0.5cm} Reference  \\ \hline \hline
 Pt & $d_{\rm Pt}$ & 2.98 + $i$ 6.37 &  Werner\cite{Werner2009}  \\
Fe$_{50}$Co$_{50}$ & $d_{\rm FeCo}$ & 3.69 + $i$ 4.53 & Werner\cite{Werner2009}   \\
MgO & 10 & 1.73 & Stephens {\it et al.}\cite{Stephens1952}    \\
SiO$_2$ & 100 & 1.45  & Malitson\cite{Malitson1965}   \\
Si & 500000 & 3.68 + $i$ 0.0054 & Green\cite{Green2008}
\\ \hline \hline
\end{tabular}
\end{table*}

\begin{figure}[t]
%\begin{center}
\includegraphics[width=10cm,keepaspectratio,clip]{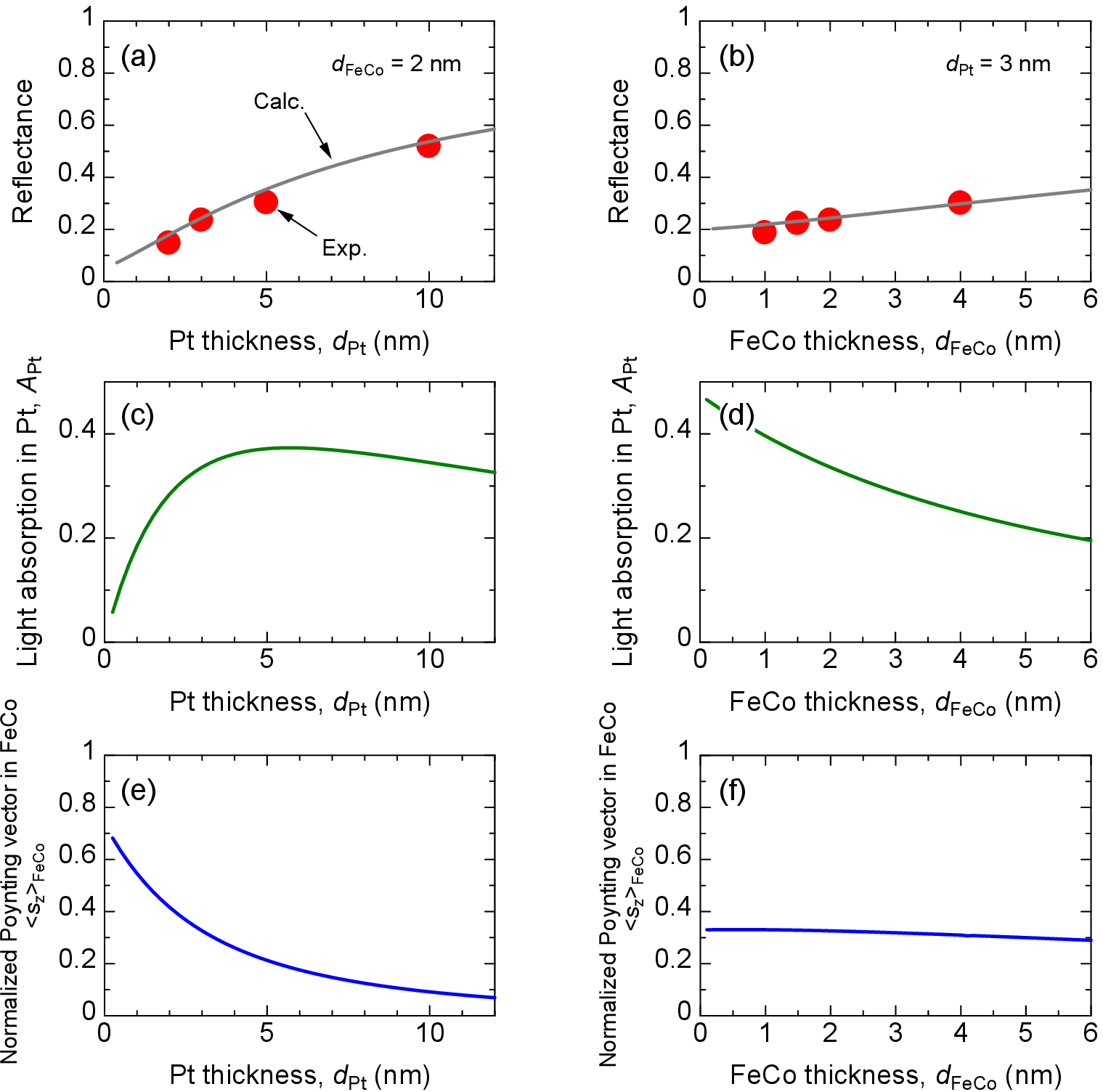}
%\end{center}
\caption{Dependence of reflectance on (a) Pt and (b) FeCo thickness. The solid curves denote the calculation results obtained via the transfer matrix method. Dependence of calculated light absorption in the Pt layer on (c) Pt and (d) FeCo thickness. Dependence of the normalized Poynting vector inside the FeCo layer on (e) Pt and (f) FeCo thickness.}
\label{f4}
\end{figure}

\clearpage

\section{Poynting vectors and light absorption}

The Poynting vectors and light absorption in s-polarization are described below. No significant difference exists between s-polarization and p-polarization since the light is incident close to the film normal. 
The normal component of the Poynting vector $s_{\rm z}$ is calculated by,
\begin{align}
s_{\rm z}(z)=\frac{ {\rm Re} \left[ \tilde{n} \cos \theta \left( E_{\rm +}^{\star }(z) + E_{\rm -}^{\star }(z)\right) \left( E_{\rm +}(z) - E_{\rm -}(z)\right) \right] }{ \cos \theta _0 } ,
\end{align} 
where $\tilde{n}$, $\theta $, $E_{\rm +}$, and $E_{\rm -}$ are the complex refractive index, incident angle, electric field for forward light, and electric field for backward light in the layer, respectively. 
$\theta _0$ is the angle of incidence from the air.
The $E_{\rm +}(z)$, $E_{\rm +}^{\star }(z)$, $E_{\rm -}(z)$, $E_{\rm -}^{\star }(z)$ can be calculated as,
\begin{align}
&E_{\rm +}(z) = E_{\rm +}(0) \exp (ik_{\rm z} z) \exp (-k^{\prime }_{\rm z} z), \\
&E_{\rm +}^{\star }(z) = E_{\rm +}^{\star }(0) \exp (-i k_{\rm z} z) \exp (-k^{\prime }_{\rm z} z), \\
&E_{\rm -}(z) = E_{\rm -}(0) \exp (-ik_{\rm z} z)\exp (k^{\prime }_{\rm z} z), \\
&E_{\rm -}^{\star }(z) = E_{\rm -}^{\star }(0) \exp (ik_{\rm z} z) \exp (k^{\prime }_{\rm z} z) ,
\end{align}
where $k_{\rm z} $ and $k^{\prime }_{\rm z} $ are real and imaginary parts of the wavenumber of light, respectively.
$E_{\pm }(0)$ is the electric field amplitude at the interface of the layer. 
The light absorption $a(z)$ in the layer is obtained by,
\begin{align}
a(z) = -\frac{ds_{\rm z}(z)}{dz}.
\end{align}
The light absorption in Pt layer shown in the main text is calculated by,
\begin{align}
A_{\rm Pt} = \int _0 ^{d_{\rm Pt}} a_{\rm Pt}(z) dz. \label{eq:APt}
\end{align}
where $a_{\rm Pt}(z)$ is the light absorption in the Pt layer.
In contrast, the average Poynting vector inside the FeCo layer $\left< s_{\rm z} \right> _{\rm FeCo}$ is calculated by,
\begin{align}
\left< s _{\rm z} \right>_{\rm FeCo} = \frac{1}{d_{\rm FeCo}}\int _0 ^{\rm d_{\rm FeCo}} s_{\rm z, FeCo}(z) dz,
\end{align}
where $s_{\rm z, FeCo}$ is a normal component of the Poynting vector inside the FeCo layer.
 The light absorption in Pt layer and average Poynting vector with different $d_{\rm Pt}$ and $d_{\rm FeCo}$ are shown in Figs. S1(c) \-- 1(f).

\clearpage

\color{black}
\section{Kerr rotation angle and magneto-optical constant}

\begin{figure}[ht]
\begin{center}
\includegraphics[width=11cm,keepaspectratio,clip]{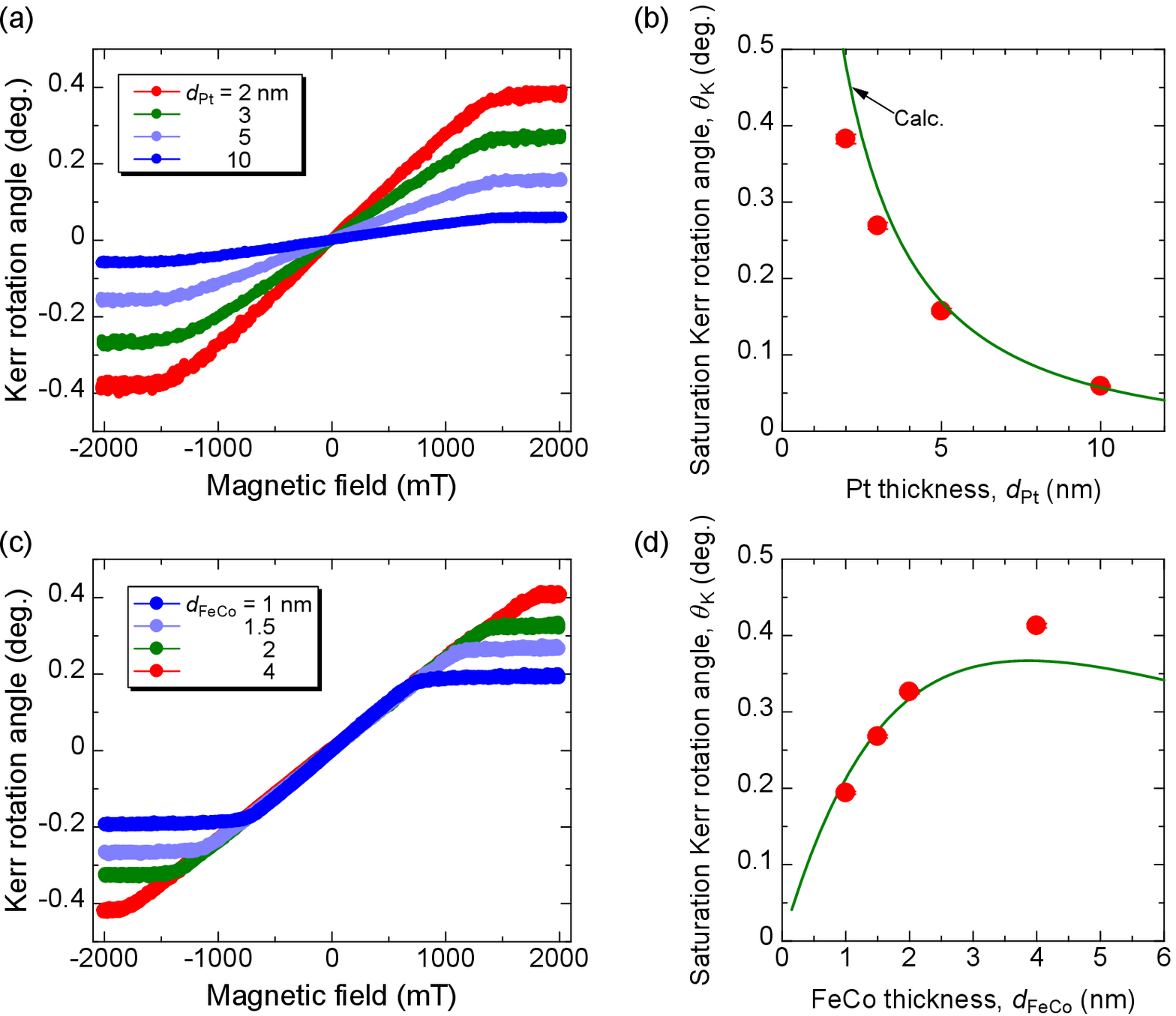}
\end{center}
\caption{(a) Kerr rotation angle plotted as a function of magnetic field with different Pt thickness. (b) Pt thickness dependence of the saturation Kerr rotation angle. (c) Kerr rotation angle plotted as a function of magnetic field with different FeCo thickness. (d) FeCo thickness dependence of the saturation Kerr rotation angle. Solid curves in (b), (d) are calculated results by using transfer matrix method. }
\label{fS_MO}
\end{figure}

Here, the Kerr rotation angle with different thicknesses and the evaluation of magneto-optical constant $Q$ in the FeCo layer are presented.
Figure S\ref{fS_MO} shows the measured Kerr rotation angle with different Pt [Fig. S\ref{fS_MO}(a), (b)] and FeCo thicknesses [Fig. S\ref{fS_MO}(c), (d)].
The dielectric tensor in the presence of the magnetization pointing along $z$-direction can be expressed as,
\begin{align}
\tilde{\varepsilon } = \varepsilon _{\rm xx} \left( 
\begin{array}{ccc}
1 & i Q m_{\rm z} & 0 \\
-i Q m_{\rm z} & 1 & 0 \\
0 & 0 & 1
\end{array}
\right) ,
\end{align}
where $\varepsilon _{\rm xx}$ and $Q$ are diagonal element of permeability and the magneto-optical constant, respectively.
The Kerr rotation angle of the multilayer stack was calculated by the transfer matrix method\cite{Zak1990, Qiu2000}.
Solid curves in Fig. \ref{fS_MO}(b), (d) are calculated results with the magneto-optical constant $Q$ = 0.032.
The $Q$ is close to the value reported previously\cite{Krebs1979}.
Here, the imaginary part of $Q$ is not considered since imaginary part is usually small and does not affect the inverse Faraday effect.  
The Faraday rotation angle $\theta _{\rm F}/d_{\rm F}$ = 4.64 m$^{-1}$ is obtained by using the relation $\theta _{\rm F}/d_{\rm F} = \pi nQ/\lambda $.

\color{black}

\clearpage

\section{In-plane torque due to inverse Faraday effect}

\begin{figure}[ht]
\begin{center}
\includegraphics[width=11cm,keepaspectratio,clip]{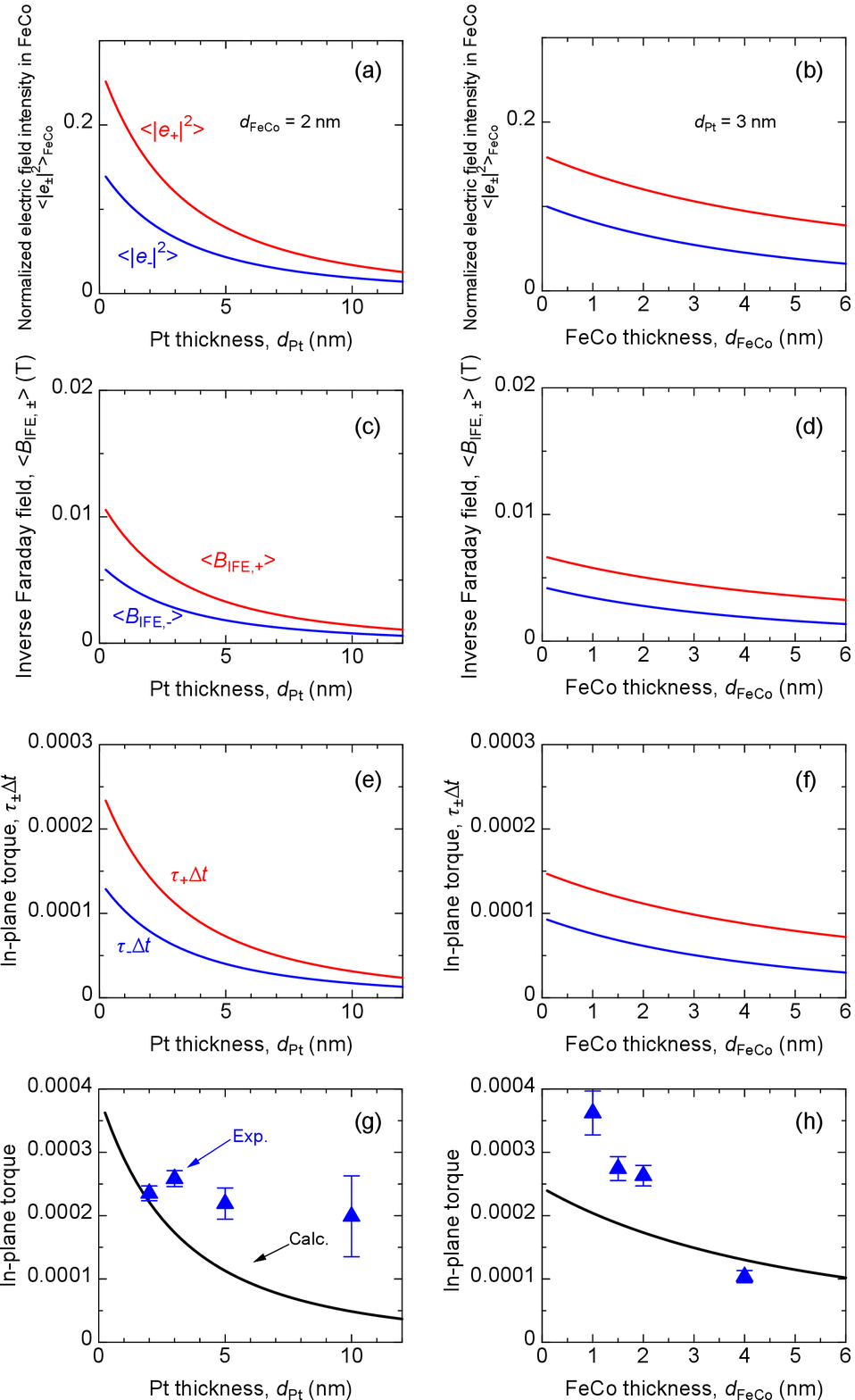}
\end{center}
\caption{Dependence of normalized electric field intensity in the FeCo layer on (a) Pt and (b) FeCo thickness. Dependence of inverse Faraday field calculated by Eq. (\ref{eq:BIFE}) on (c) Pt and (d) FeCo thickness. Dependence of in-plane torque, $\tau _{\pm }\Delta t$ calculated using Eq. (\ref{eq:tau}) on (e) Pt and (f) FeCo thickness. Dependence of calculated in-plane torque due to inverse Faraday effect and experimental in-plane torque on (g) Pt and (h) FeCo thickness. }
\label{fS_IFE}
\end{figure}

Here, the in-plane torque due to the inverse Faraday effect is calculated.
Based on Refs. \cite{Pershan1965, Kirilyuk2010, Choi2019}, the magnetic field due to the inverse Faraday effect in ferromagnetic film is calculated as,
\begin{align}
B_{\rm IFE}&=\frac{\varepsilon _0 n \lambda \theta _{\rm F}|E|^2}{2\pi M_{\rm s}d_{\rm F}},
\end{align}
where $n$, $\lambda $, $\theta _{\rm F}$, $d_{\rm F}$ and $M_{\rm s}$ are the refractive index, wavelength, Faraday rotation angle, ferromagnetic layer thickness, and saturation magnetization, respectively.
By using the relation $F_{\rm p}/\Delta t = c\varepsilon _0 |E_0|^2/2$, the following equation for Inverse Faraday field in the ferromagnetic layer is derived:
\begin{align}
B_{\rm IFE, \pm}(z) = \frac{F_{\rm p}n\lambda }{\pi c\Delta tM_{\rm s}}\cdot \frac{\theta _{\rm F}}{d_{\rm F}}|e_{\pm }(z)|^2, \notag
\end{align}
where $c$, $\Delta t$, and $e_{\pm }$ are speed of light, pulse duration, and normalized light intensity for $\pm $ light, respectively.
Here, the relation $E_{\pm }(z)=E_0 e_{\pm }(z)$ is used.
The average $B_{\rm IFE, \pm}$ is obtained by taking the average of light intensities inside the FeCo layer as,
\begin{align}
&\left< |e_{\pm }|^2 \right> = \frac{1}{d_{\rm FeCo}}\int ^{d_{\rm FeCo}}_{0} |e_{\pm } (z)|^2 dz, \notag \\
&\left< B_{\rm IFE, \pm} \right> = \frac{F_{\rm p}n\lambda }{\pi c\Delta tM_{\rm s}} \cdot \frac{\theta _{\rm F}}{d_{\rm F}} \left< |e_{\pm }|^2 \right> . \label{eq:BIFE}
\end{align}
$\left< |e_{\pm }|^2 \right>$ and $\left< B_{\rm IFE, \pm }\right>$ with different $d_{\rm Pt}$ and $d_{\rm FeCo}$ are shown in Figs. S\ref{fS_IFE}(a) \-- (d).
The time integrated in-plane torque due to $\left< B_{\rm IFE, \pm }\right> $ is calculated as,
\begin{align}
\tau _{\pm }\Delta t = \gamma \left< B_{\rm IFE, \pm}\right> \Delta t.  \label{eq:tau}
\end{align}
$\tau _{\pm }\Delta t$ with different $d_{\rm Pt}$ and $d_{\rm FeCo}$ calculated are shown in Figs. S\ref{fS_IFE}(e) and S\ref{fS_IFE}(f).
Here, the parameters used in the calculation are $M_{\rm s}$ = 2.4 MA/m, \color{black}$\theta _{\rm F}/d_{\rm F}$ = 4.6 $\times $ 10$^5$ m$^{-1}$\color{black}, $n$ = 3.69.
 
The experimental in-plane torque and calculated \color{black}$(\tau _{+} + \tau _{-})\Delta t$ \color{black} are shown in Figs. S\ref{fS_IFE}(g) and S\ref{fS_IFE}(h).
\color{black}Note that $B_{\rm IFE}$ for $\pm $ light has same signs whereas the helicity of reflected light is reversed owing to the reversal of the coordinate\cite{Choi2019}.
The significant increase of the experimentally evaluated in-plane torque with decreasing FeCo thickness cannot be explained by the inverse Faraday effect inside the FeCo layer. 
\color{black}

\clearpage

\section{Angular momentum conversion efficiency}

The optical orientation of the Pt layer can induce spin in the Pt layer and transfer its angular momentum to the FeCo layer via the spin-transfer torque effect\cite{Choi2020}.Here, the angular momentum conversion efficiency from light to spin in the Pt layer is evaluated.

\begin{figure}[ht]
\begin{center}
\includegraphics[width=10cm,keepaspectratio,clip]{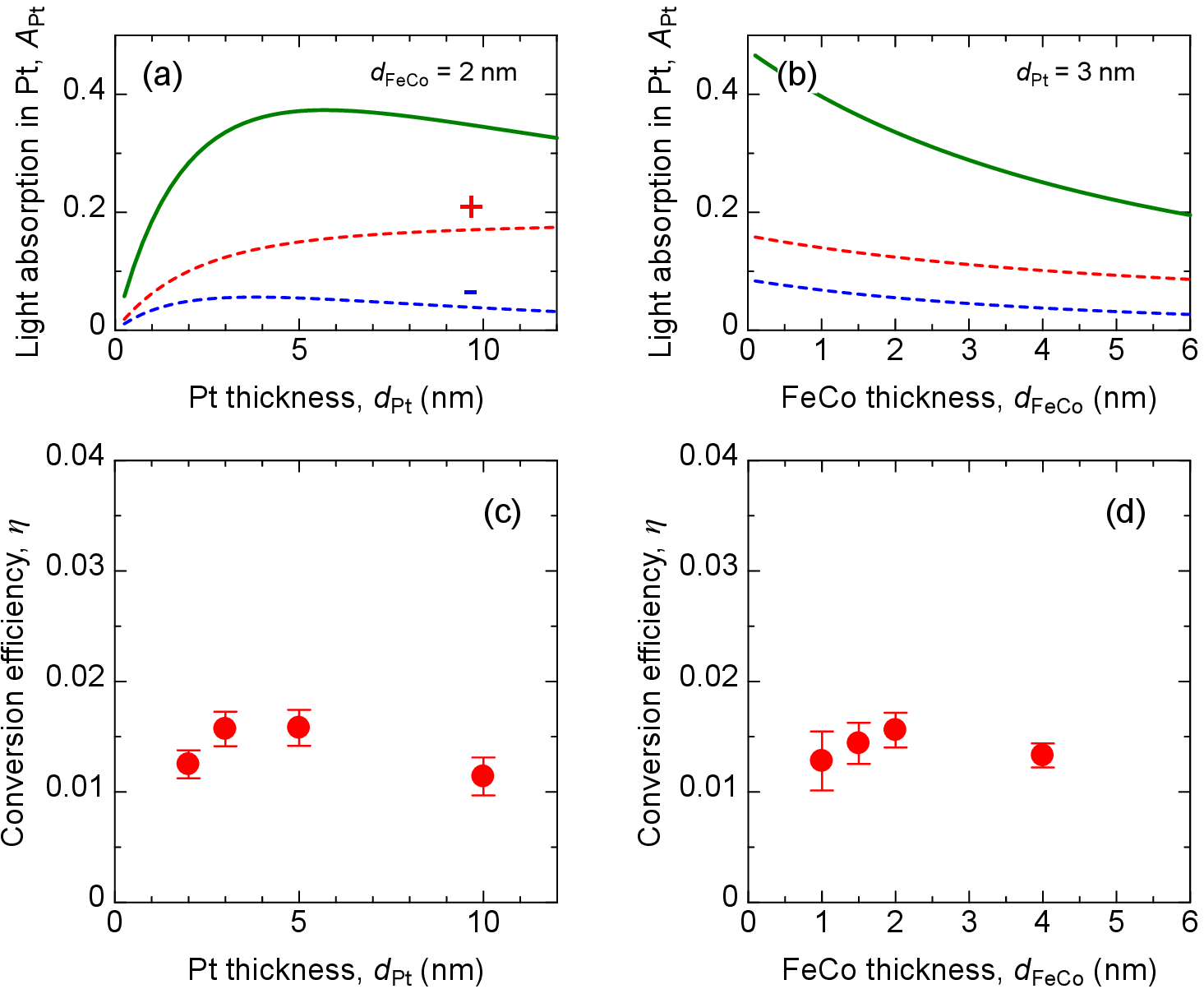}
\end{center}
\caption{Dependence of light absorption in the Pt layer on (a) Pt and (b) FeCo thickness. The light absorption of $\pm $ light are shown as red and blue broken curves. Dependence of the conversion efficiency evaluated by Eq.(S12) on (c) Pt and (d) FeCo thickness.}
\label{fS_eta}
\end{figure}

\begin{figure}[ht]
\begin{center}
\includegraphics[width=6cm,keepaspectratio,clip]{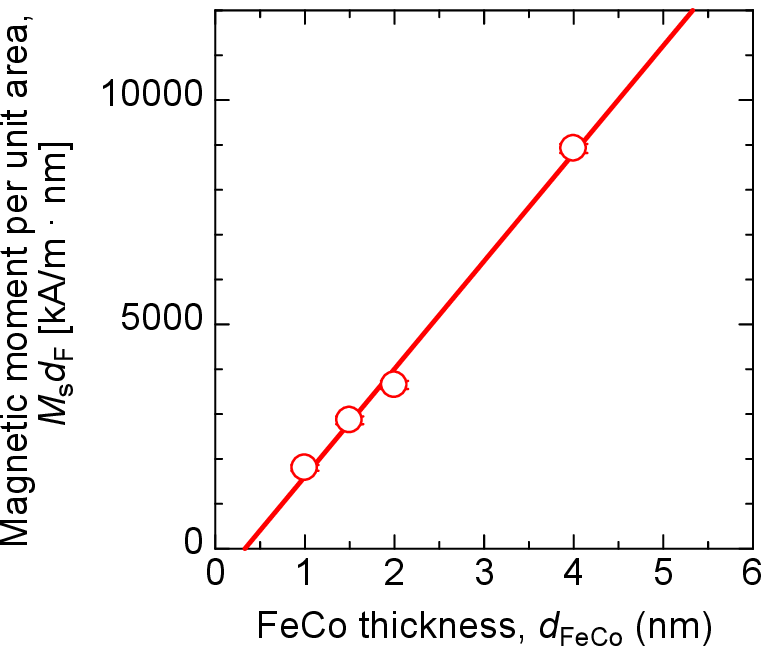}
\end{center}
\caption{Magnetic moment per unit area measured by vibrating sample magnetometer plotted as a function of $d_{\rm FeCo}$.}
\label{fS_Msd}
\end{figure}

The absorbed photon number for $\pm $ light per unit area $n_{\rm p, \pm}$ in the Pt layer can be calculated as follows:
\begin{align}
n_{\rm p, \pm } = \frac{A_{{\rm Pt}, \pm }F_{\rm p}}{\hbar \omega }, \notag 
\end{align}
where $\hbar \omega $ is a photon energy of 1.55 eV. 
$F_{\rm p}$ is pump fluence used in the experiment.
$A_{{\rm p}, \pm}$ is the light absorption for $\pm $ light, which is calculated using Eqs. (S1)--(\ref{eq:APt}) as,
\begin{align}
&A_{{\rm Pt}, \pm} = \int _0 ^{\rm Pt} a_{\rm Pt, \pm}(z) dz, \notag \\
&a_{\rm Pt, \pm}(z)=\mp \frac{d}{dz} \left( \frac{{\rm Re} \left( \tilde{n} \cos \theta E_{\pm }^{\star }(z) E_{\pm }(z) \right) }{\cos \theta _0} \right).
\end{align}
$d_{\rm Pt}$ and $d_{\rm FeCo}$ dependences of $A_{{\rm Pt}, \pm }$ are shown as broken curves in Fig. S\ref{fS_eta}(a) and S\ref{fS_eta}(b).
The spin-angular momentum of the absorbed photon per unit area $s_{\rm p}$ can be calculated as follows: \color{black}
\begin{align}
s_{\rm p} = \hbar \left( n_{{\rm p}, +} + n_{{\rm p}, -} \right) . \label{eq:sp}
\end{align}
\color{black}
In contrast, the angular momentum corresponding to the change in magnetization of the FeCo layer induced by the spin-transfer torque can be calculated as,
\begin{align}
s_{\rm m} = \frac{\delta m_{\rm z} M_{\rm s}d_{\rm F} }{\gamma }, \label{eq:sm}
\end{align}
where $\delta m_{\rm z}$, $M_{\rm s}d_{\rm F}$, and $\gamma $ are change in out-of-plane component of the normalized magnetization induced by the spin-transfer torque, magnetic moment per unit area, and the gyromagnetic ratio, respectively.
$M_{\rm s}d_{\rm F}$ was measured by vibrating sample magnetometer as shown in Fig. S\ref{fS_Msd}. 
$M_{\rm s}d_{\rm F}$ can be expressed as, $M_{\rm s}d_{\rm F} = M_{\rm s}^{\star }(d_{\rm FeCo} -d^{\star })$.
$M_{\rm s}^{\star }$ and $d^{\star }$ are evaluated to be 2.40 $\pm $ 0.13 MA/m and 0.33 $\pm $ 0.13 nm, respectively.

By using Eqs. (\ref{eq:sp}) and (\ref{eq:sm}), we can calculate the angular momentum conversion efficiency $\eta $ as,
\begin{align}
\eta = \frac{s_{\rm m}}{s_{\rm p}}. \label{eq:eta}
\end{align}
Figures S\ref{fS_eta}(c) and (d) show $\eta $ plotted as functions of Pt and FeCo thickness. 
It was found that $\eta $ is 0.01 $\sim $ 0.02 and almost constant with $d_{\rm Pt}$ and $d_{\rm FeCo}$.

\clearpage

\color{black}
\section{Calculation of the interface spin and the in-plane torque generated by the circularly polarized light in the presence of Rashba spin-orbit coupling}

In this section, calculation of interface spin in the presence of Rashba spin-orbit coupling is presented.
Rashba spin-orbit coupling is given by Eq. (5) shown in the main text.
The magnetic field ${\bf H}_{\rm R}$ orthogonal to both the electron momentum vector ${\bf p}$ and the axis of the inversion symmetry breaking ${\bf z}$ can be generated by the Rashba spin-orbit coupling as,
\begin{align}
{\bf H}_{\rm R} = \frac{2\alpha _{\rm R}}{\gamma \hbar ^2} \left( {\bf p} \times {\bf z} \right) . \label{eq:vecHr}
\end{align}
The equation of motion of the electron is approximated classically as, $d{\bf p}/dt = -e{\bf E} -{\bf p}/\tau $,
where, $e$ and $\tau $ are elementary charge and electron kinetic momentum lifetime, respectively.
The circularly polarized electric field is described as, ${\bf E} = E_0 \exp (i\omega t)(i {\bf x} + {\bf y})$.
The solution is obtained as, ${\bf p} = -{\bf E}e\tau \exp (i\theta )/\sqrt{1+(\omega \tau )^2}$, where $\theta = -\tan ^{-1} (\omega \tau )$.
Therefore, ${\bf H}_{\rm R}$ in Eq. (\ref{eq:vecHr}) can be expressed as,
\begin{align}
{\bf H}_{\rm R} = h_{\rm R} \left( \cos (\omega t + \theta ) {\bf x} + \sin (\omega t + \theta ) {\bf y}\right) . \label{eq:vecHr2}
\end{align}
where, $h_{\rm R}$ is given by,
\begin{align}
h_{\rm R} &= -\frac{2\alpha _{\rm R}eE_0 \tau }{\gamma \hbar ^2\sqrt{1+(\omega \tau )^2}} \notag \\
&\approx -\frac{2\alpha _{\rm R}eE_0 }{\gamma \hbar ^2 \omega }. \label{eq:hr}
\end{align}
Here, a approximation $(\omega \tau )^2 \gg 1$ is used.
Therefore, circularly polarized light with the presence of Rashba spin-orbit coupling can generate rotating effective magnetic field with the strength $h_{\rm R}$.

Next, the equation of spin in the 2-dimentional electron system is discussed based on the Bloch equation.
The magnetization ${\bf M}_{\rm R}$ in the  Bloch equation with the rotating reference frame $({\bf x}^{\prime }= {\bf x} \cos \omega t + {\bf y} \sin \omega t, {\bf y}^{\prime }= -{\bf x}\sin \omega t + {\bf y} \cos \omega t)$ is given by,
\begin{align}
\frac{d{\bf M}_{\rm R}}{dt} = -\gamma {\bf M}_{\rm R} \times {\bf H} - \frac{1}{\tau _{\rm s}} \left( {\bf M}_{\rm R} - \chi _{\rm P} {\bf H}\right) - \omega {\bf M}_{\rm R} \times {\bf z},  \label{eq:Bloch}
\end{align}
where $\chi _{\rm P}$ and $\tau _{\rm s}$ are the Pauli susceptibility and spin lifetime, respectively.
The third term can be considered as an additional magnetic field $(\omega {\bf z} /\gamma)$ due to the transformation to the rotating frame.
The static solution of Eq. (\ref{eq:Bloch}) is obtained by sustituting $d{\bf M}_{\rm R}/dt = 0 $.
The matrix form of the equation can be expressed as,
\begin{align}
\left(
\begin{array}{ccc}
1 & \omega \tau _{\rm s} & 0 \\
-\omega \tau _{\rm s} & 1 & \gamma \tau _{\rm s} h_{\rm R} \\
0 & -\gamma \tau _{\rm s} h_{\rm R} & 1 
\end{array}
\right) \left(
\begin{array}{c}
M_{\rm R, x^{\prime }} \\ M_{\rm R, y^{\prime }} \\ M_{\rm R, z}
\end{array}
\right) = \chi _{\rm P} \left( 
\begin{array}{c}
h_{\rm R} \\ 0 \\ 0
\end{array}
\right) .
\end{align}
The solution of $M_{\rm R, z}$ is obtained as,
\begin{align}
M_{\rm R, z} &= \frac{\gamma \omega \tau _{\rm s}^2 h_{\rm R}^2 }{1+(\gamma \tau _{\rm s} h_{\rm R})^2+(\omega \tau _{\rm s})^2}\chi _{\rm P} \notag \\
&\approx \frac{\gamma h_{\rm R}^2}{\omega }\chi _{\rm P}. \label{eq:Mz}
\end{align}
Here, approximations $1 \ll (\omega \tau _{\rm s})^2$, $h_{\rm R}^2 \ll (\omega /\gamma )^2$ are used. By using Eqs. (\ref{eq:hr}) and (\ref{eq:Mz}), one can obtain following expression for the z-component spin $s_{\rm R}$,
\begin{align}
s_{\rm R, z} = -\frac{M_{\rm R, z}}{\gamma } = \chi _{\rm P}\frac{4\alpha _{\rm R}^2 e^2}{\gamma ^2\hbar \omega ^3}E_0^2 . \label{eq:sR}
\end{align}
Here, $\chi _{\rm P}$ for the 2-dimentional electron can be given by,
\begin{align}
\chi _{\rm P} &= \frac{2\mu _{\rm B}^2 D(\epsilon _{\rm F})}{S} \notag \\
&=\frac{\mu _{\rm B}^2 m_0}{\pi \hbar }, \label{eq:chiP}
\end{align}
where, $D(\epsilon _{\rm F})$, $\mu _{\rm B}$, $m_0$ and $S$ are density of states at the Fermi level, the Bohr magneton, electron mass, and surface area, respectively.
Here, the relation $D(\epsilon_{\rm F}) = S m_0 /(2\pi \hbar ^2)$ is used for the 2-dimensional electron system.
Finally, using Eqs. (\ref{eq:sR}) and (\ref{eq:chiP}) with the relation $\gamma \approx 2\mu _{\rm B}/\hbar $, one can obtain a following equation,
\begin{align}
s_{\rm R, z} = \alpha _{\rm R}^2 \cdot \frac{m_0 e^2}{\pi \hbar ^4 \omega ^3} E_0^2 . \label{eq:sRz_cla}
\end{align}

In addition to the above expression for the interface spin, we describe the equation from Edelstein\cite{Edelstein1997} as well.
From Eqs. (7) and (8) in \cite{Edelstein1997}, the interface spin generated by the circularly polarized light is described by,
\begin{align}
{\bf s}_{\rm R} = i {\bf z} \left( {\bf z} \cdot {\boldsymbol  \epsilon} \times {\boldsymbol  \epsilon}^{\star }\right) K E_0 ^2 ,
\end{align} 
where, ${\boldsymbol  \epsilon}$ is a unit vector of ${\bf E}$. $K$ is given by\cite{Edelstein1997},
\begin{align}
K= \frac{e^2}{2\pi ^2 \epsilon_{\rm F} \omega ^3 \tau ^2}\cdot \frac{\tau /\tau _{\rm s. o.}}{\tau / \tau _{\rm s. o.} + \tau /\tau _{\rm s}},
\end{align}
where, $\tau _{\rm s. o.}$ is spin lifetime due to Dyakonov-Perel process ($1/\tau _{\rm s. o.} = \alpha _{\rm R}^2 k_{\rm F}^2 \tau /\hbar ^2 $, where $k_{\rm F}$ is the Fermi wavenumber). 
The above equation can be simplified by assuming $1/\tau _{\rm s. o.} \ll 1/\tau _{\rm s}$ as,
\begin{align}
K= \alpha _{\rm R}^2 \cdot \frac{m_0 e^2 }{\pi ^2 \hbar ^4 \omega ^3}\cdot \frac{\tau _{\rm s}}{\tau }. \label{eq:K}
\end{align}
By comparing Eqs. (\ref{eq:sRz_cla}) and (\ref{eq:K}), the difference is a factor $\pi \tau /\tau _{\rm s} $. 
Based on above two expressions, we use the formula of the interface spin generated by the circularly polarized light with the presence of the Rashba spin-orbit coupling $\sim \alpha _{\rm R}^2 m_0 e^2 E_0^2 /(\pi \hbar ^4 \omega ^3) $.

The $sd$ exchange coupling between the conduction electron spin and the magnetization can induce the torque on the magnetization.
The exchange coupling per unit area between electron spin ${\bf s}_{\rm R}$ and the magnetization ${\bf m}$ can be expressed as,
\begin{align}
\mathcal{H}_{\rm sd} = \frac{2J_{\rm sd}}{\hbar } {\bf s_{\rm R} } \cdot {\bf m},
\end{align}
where $J_{\rm sd}$ and ${\bf m}$ are $sd$ exchange coupling constant and unit vector of the local magnetization.
The magnetic field $H_{\rm sd}$ due to $sd$ exchange coupling acting on ${\bf m}$ can be calculated as,
\begin{align}
H_{\rm sd} &= -\frac{1}{M_{\rm s}d_{\rm F}}\cdot \frac{\partial \mathcal{H}_{\rm sd}}{\partial {\bf m}} \notag \\
&= -\frac{2J_{\rm sd}}{\hbar M_{\rm s}d_{\rm F}}{\bf s}_{\rm R},
\end{align}
where $M_{\rm s}$ and $d_{\rm F}$ are saturation magnetization and thickness of the ferromagnetic layer, respectively.
Then, one can obtain the in-plane torque $\tau _{\parallel } \Delta t$ due to $H_{\rm sd}$ as,
\begin{align}
\tau _{\parallel }\Delta t &= \gamma H_{\rm sd} \Delta t \notag \\
&= \alpha _{\rm R}^2 \cdot \frac{2 J_{\rm sd} \gamma \Delta t  }{M_{\rm s} d_{\rm F}} \cdot \frac{m_0 e^2}{\pi \hbar ^5 \omega ^3}E_0^2 \notag \\
&= 4\alpha _{\rm R}^2\cdot \frac{ \gamma J_{\rm sd}  }{M_{\rm s} d_{\rm F}} \cdot \frac{m_0 e^2 F_p}{\pi \hbar ^5 \omega ^3 c \varepsilon _0}. \label{eq:torque}
\end{align} 
Here, the relation $F_{\rm p}/\Delta t = c\varepsilon _0 E_0^2 /2$ was used. $c$ and $\varepsilon _0$ are speed of the light and electric constant, respectively.
Finally, the in-plane torque in the unit of the angular momentum per unit area is obtained as,
\begin{align}
\frac{M_{\rm s}d_{\rm F}\tau _{\parallel }\Delta t}{\gamma } = 4\alpha _{\rm R}^2\cdot J_{\rm sd} \cdot \frac{  m_0 e^2F_{\rm p}}{\pi \hbar ^5 \omega ^3 c \varepsilon _0}. \label{eq:am}
\end{align}
When we use the typical parameters $\alpha _{\rm R} = 10 ^{-10} $ ${\rm eV} \cdot {\rm m}$\cite{Miron2010, Ast2007}, $J_{\rm sd}$ = 100 meV\cite{Barreteau2004}, the calculated angular momentum [Eq. (\ref{eq:am})] is $2.3\times 10^{-18}$ [${\rm J \cdot s \cdot m}^{-2}$], which is same order of magnitude with the experimentally evaluated value shown in the main text.
In addition, when $M_{\rm s}$ = 2 MA/m and $d_{\rm F}$ = 2 nm are used, the in-plane torque [Eq. (\ref{eq:torque})] is  1.0 $\times $ 10$^{-4}$, same order of magnitude with the observed in-plane torque. 
 
\color{black}

\clearpage

\section{Thickness dependence of the phase analysis results}

\begin{figure}[ht]
\begin{center}
\includegraphics[width=14cm,keepaspectratio,clip]{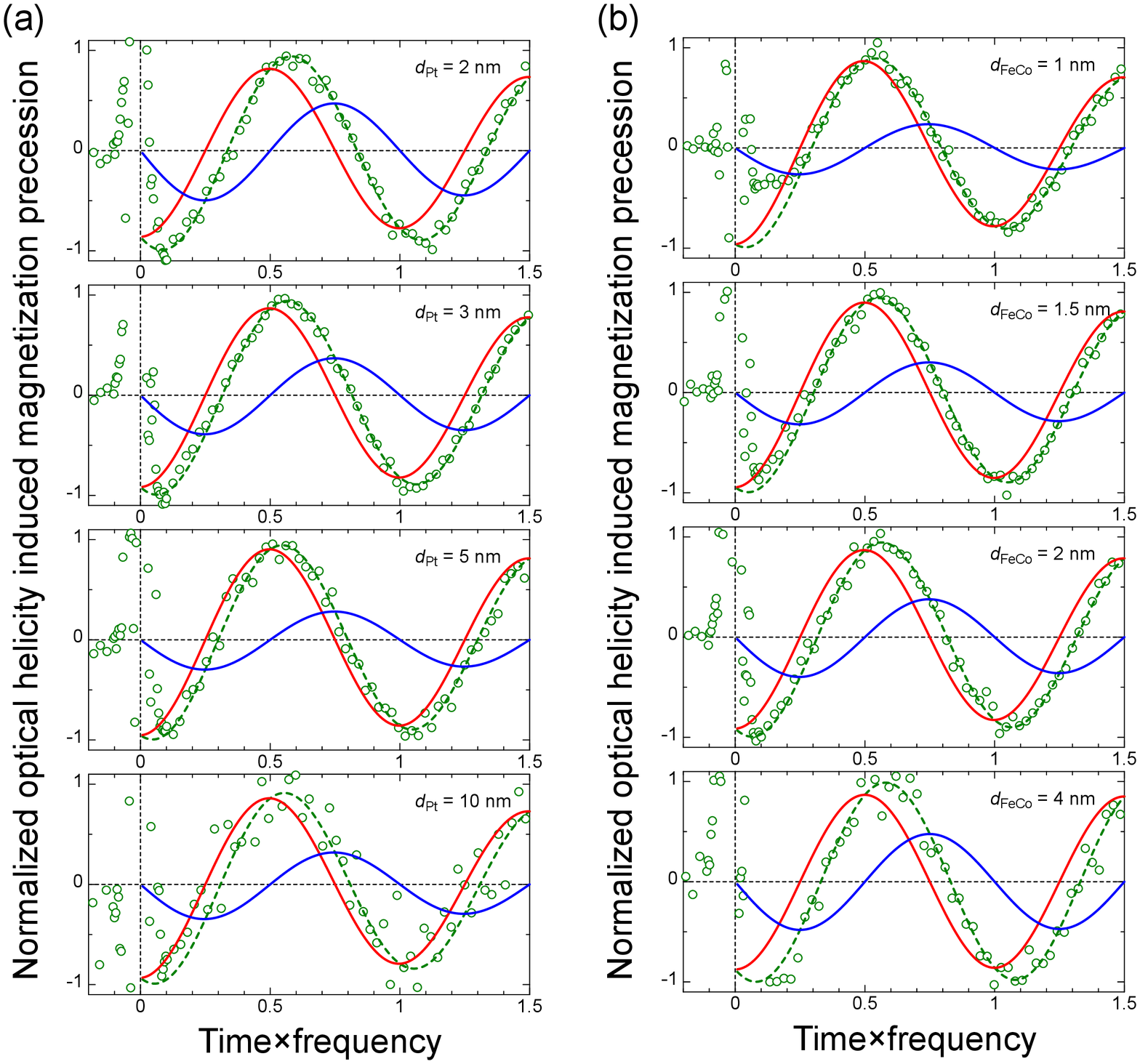}
\end{center}
\caption{ Dependence of normalized optical helicity induced magnetization precession as a function of the cycle (time normalized by the frequency) of the precession on (a) Pt and (b) FeCo thickness. The signals were decomposed into cosine (red) and sine (blue) components. }
\label{fS_phase1}
\end{figure}

\begin{figure}[ht]
\begin{center}
\includegraphics[width=14cm,keepaspectratio,clip]{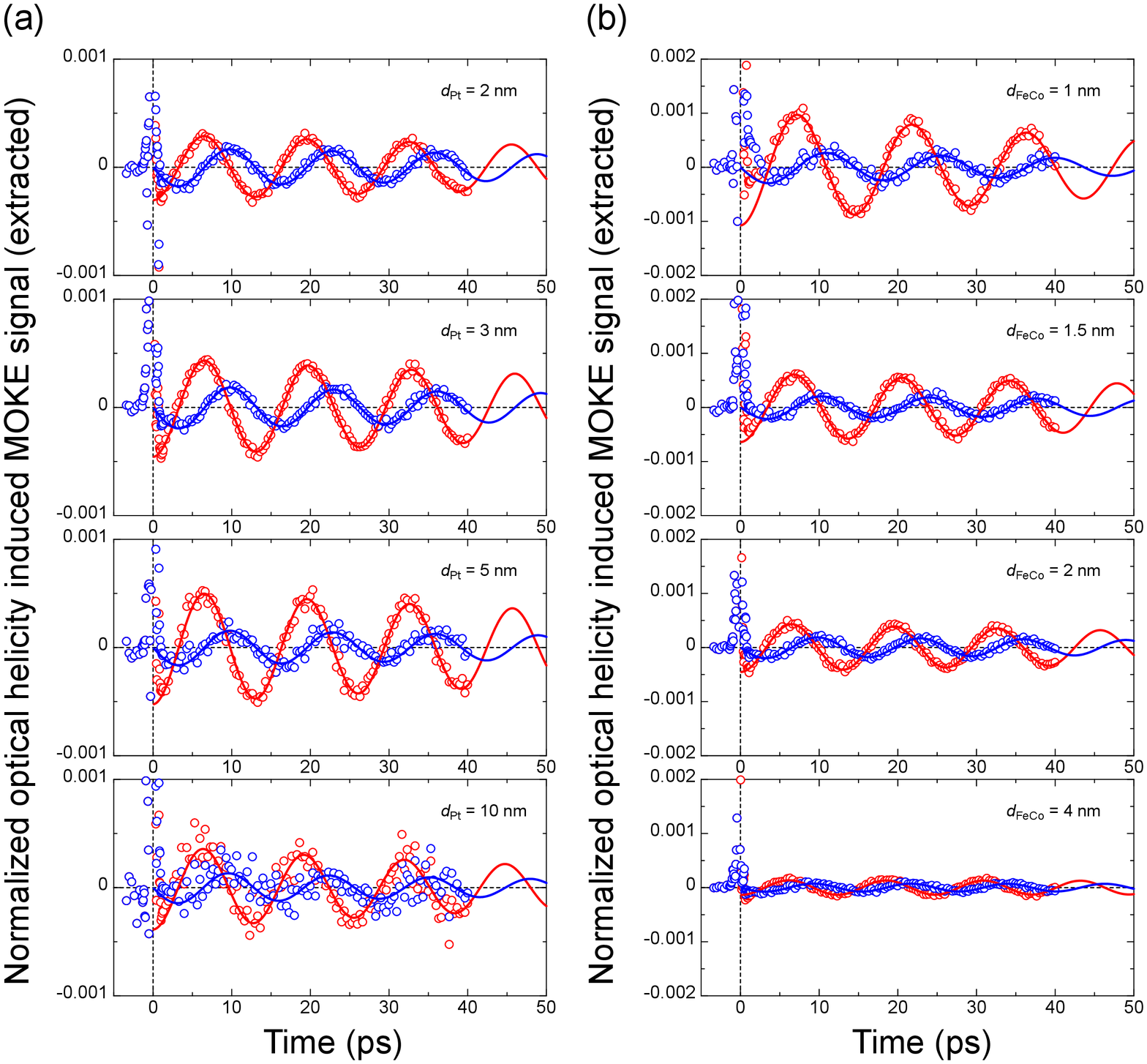}
\end{center}
\caption{Dependence of normalized MOKE signal on (a) Pt and (b) FeCo thickness. The two components, cosine (red) and sine (blue), are extracted by fitting calculations.}
\label{fS_phase2}
\end{figure}

Figure S\ref{fS_phase1} shows the thickness dependence of the optical helicity induced magnetization precession. The data are obtained by taking the difference between signals measured with LCP and RCP lights. The phase of the magnetization precession is determined by the fitting calculation and is decomposed into cosine and sine components, denoted by red and blue solid curves in Fig. S\ref{fS_phase1}, which respectively correspond to the out-of-plane and in-plane torque. Figure S\ref{fS_phase2} shows extracted cosine and sine components by fitting calculations with different $d_{\rm Pt}$ and $d_{\rm FeCo}$.

\clearpage

\section{Optical helicity induced magnetization precession with changing symmetry of the stacking structure}

\begin{figure}[ht]
\begin{center}
\includegraphics[width=16cm,keepaspectratio,clip]{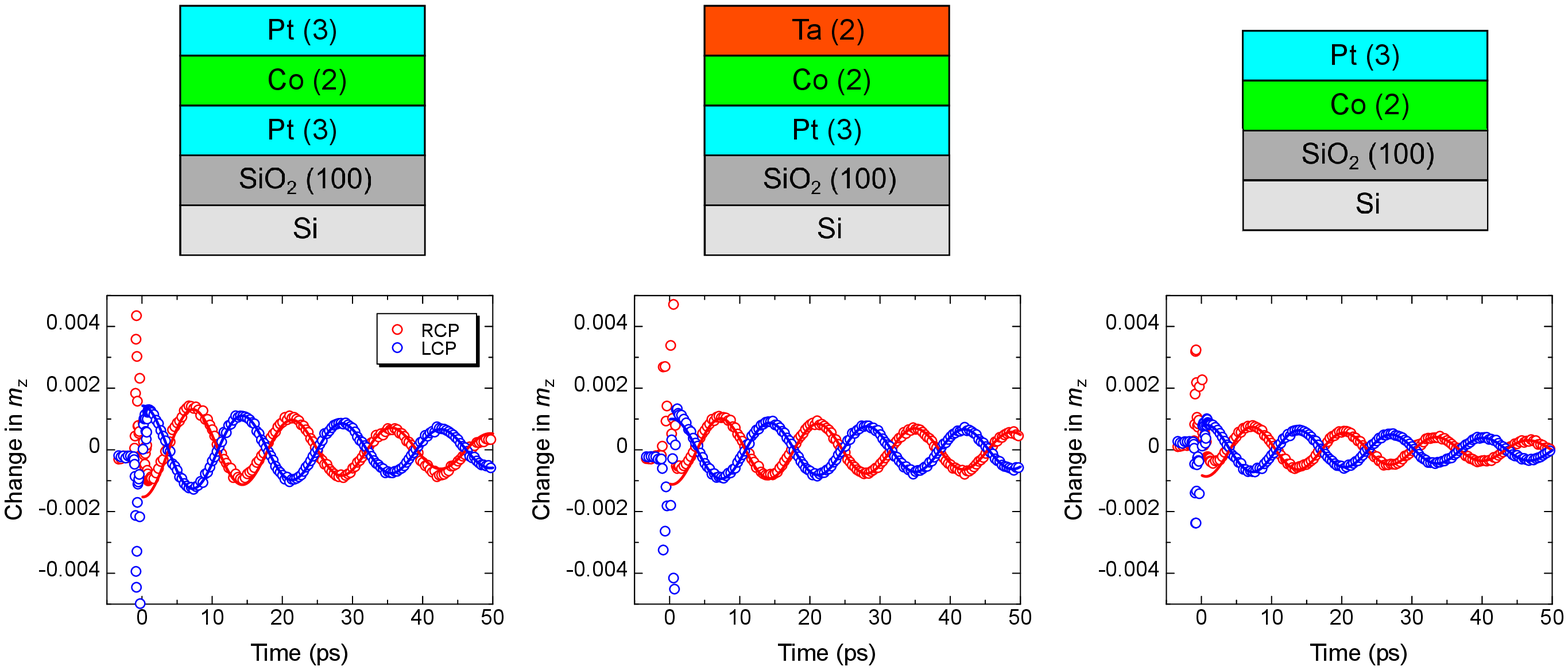}
\end{center}
\caption{(a) Circularly polarized laser induced magnetization precession with different optical helicities in three different stacking structures.}
\label{fS_CoPt}
\end{figure}

\begin{figure}[ht]
\begin{center}
\includegraphics[width=14cm,keepaspectratio,clip]{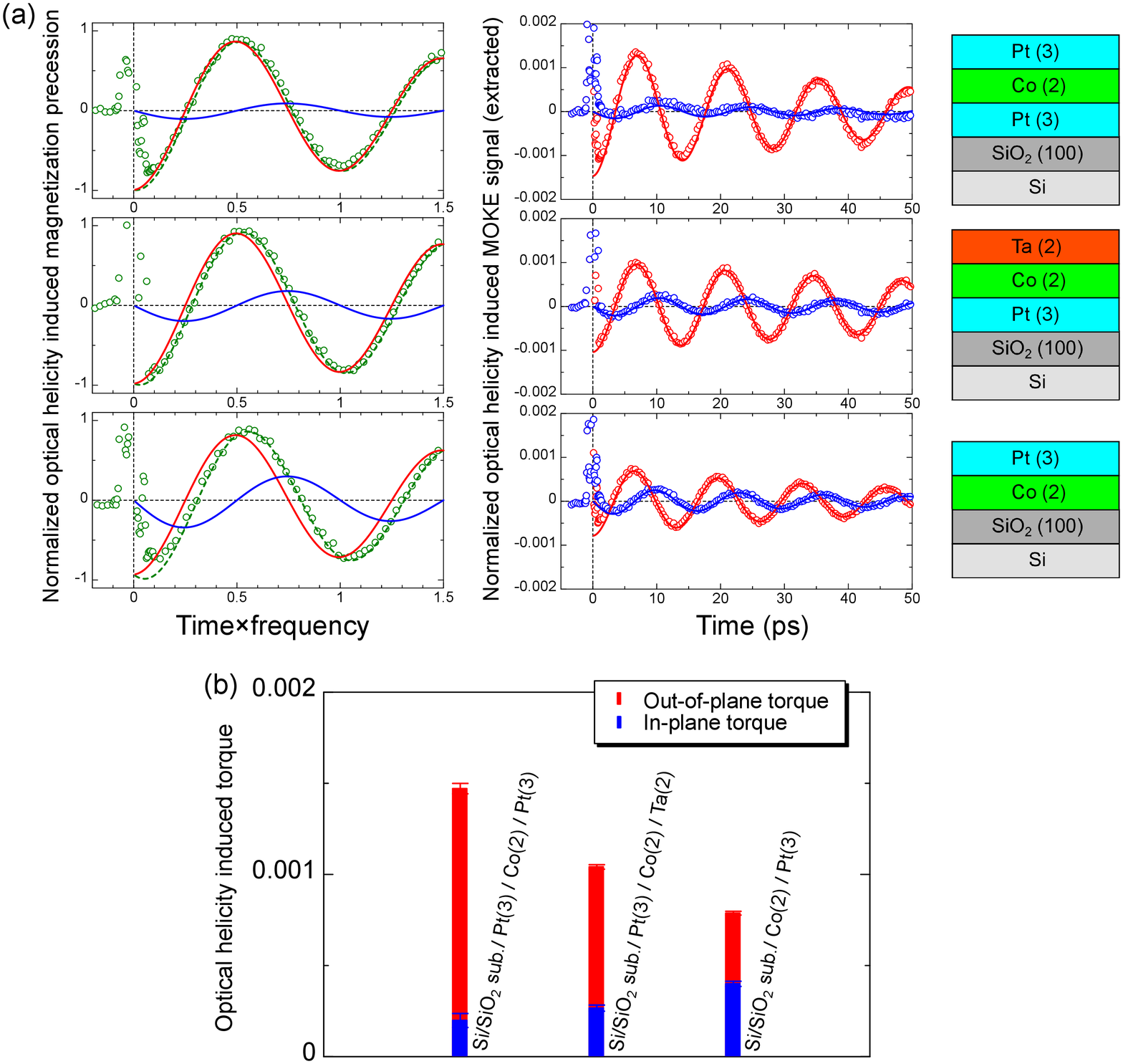}
\end{center}
\caption{(a) Optical helicity induced magnetization precession in three different stacking structures. (b) Out-of-plane and in-plane torques extracted from the phase analysis plotted for different stacking structures.}
\label{fS_CoPt2}
\end{figure}

To study the effect of the interface in the stacking structure, circularly polarized laser induced magnetization precession in the following three stacking structures are measured,
\begin{enumerate}
\item Si / SiO$_2$ sub. / Pt(3) / Co(2) / Pt(3). ``Pt/Co/Pt''
\item Si / SiO$_2$ sub. / Pt(3) / Co(2) / Ta(2). ``Pt/Co/Ta''
\item Si / SiO$_2$ sub. / Co(2) / Pt(3). ``Co/Pt''
\end{enumerate}
Here, we used $F_{\rm p}$ = 9.5 J/m$^2$.
Figure S\ref{fS_CoPt} shows circularly polarized laser induced magnetization precession with different optical helicities in the three samples listed above.
Figure S\ref{fS_CoPt2}(a) shows the optical helicity induced magnetization precession signal by taking the difference between signals measured with LCP and RCP lights. The red and blue solid curves denote the cosine and sine components of the magnetization precession extracted by the fitting calculation.
Figure S\ref{fS_CoPt2}(b) shows the overall out-of-plane and in-plane torque extracted from the observed magnetization precession in the three different samples. A large out-of-plane torque is observed for the ``Pt/Co/Pt'' sample, which is due to spin-transfer torque from both the top and bottom Pt layers. However, in-plane torque is increased when a structure with broken inversion symmetry is used, i.e., ``Co/Pt'' sample.

\begin{figure}[ht]
\begin{center}
\includegraphics[width=10.5cm,keepaspectratio,clip]{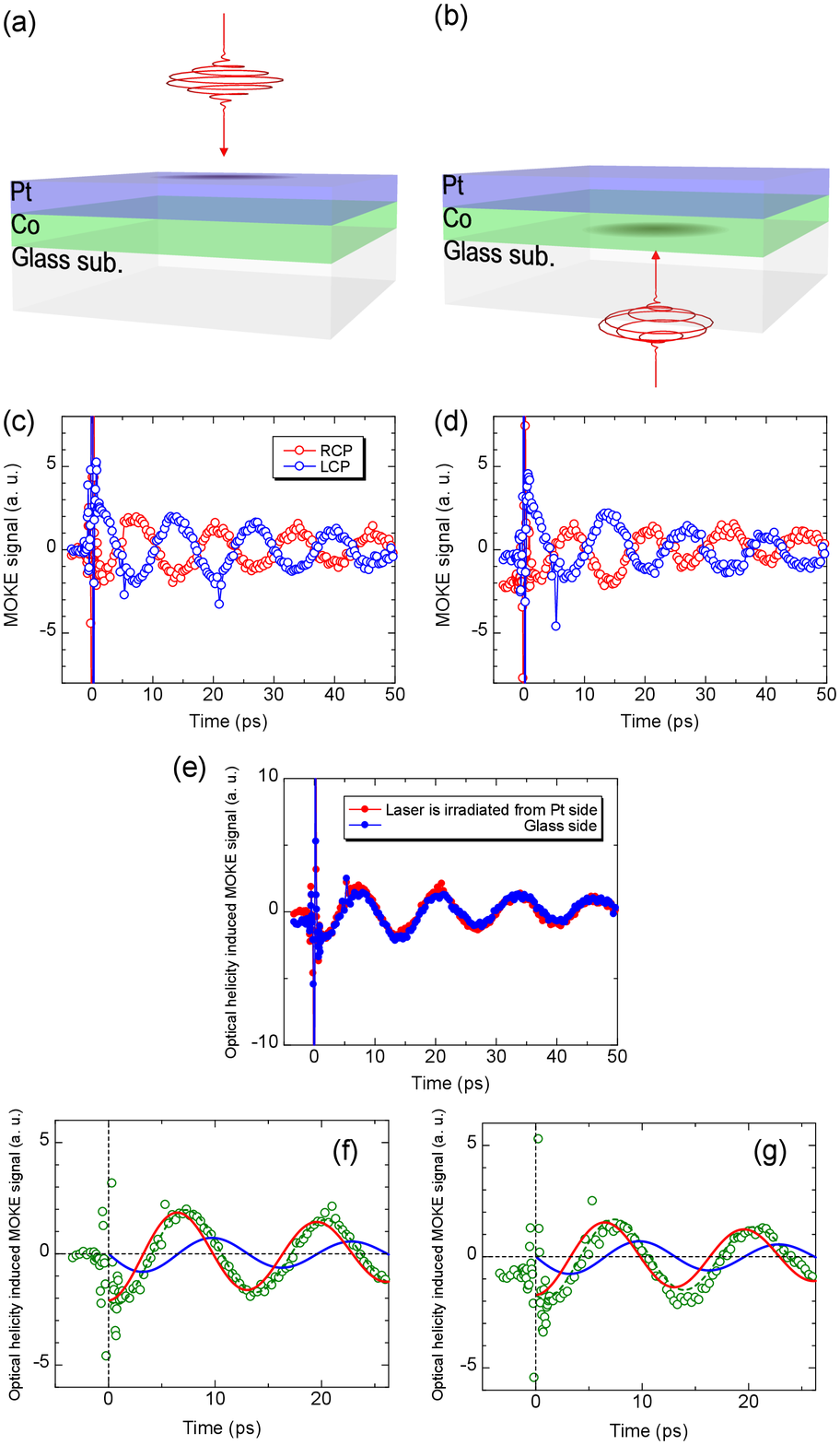}
\end{center}
\caption{Optical helicity induced magnetization precession with different experimental configurations. Circularly polarized light pulse is irradiated from (a) Pt and (b) glass substrate sides. The circularly polarized laser pulse induced magnetization precession with different optical helicities when laser is irradiated from (c) Pt and (d) glass substrate sides. (e) Comparison of optical helicity-induced magnetization precession with different experimental configurations. Analysis of the phase of magnetization precession when the laser is irradiated from (f) Pt and (g) glass substrate sides.}
\label{fS_GlassCoPt}
\end{figure}

Figure S\ref{fS_GlassCoPt} shows the experimental results when the configuration is reversed using a sample deposited on a glass substrate.
Figure S\ref{fS_GlassCoPt}(a) and S\ref{fS_GlassCoPt}(b) show two experimental configurations where circularly polarized light pulses are irradiated on a glass sub./ Co / Pt from (a) Pt and (b) glass substrate side. 
Figure S\ref{fS_GlassCoPt}(c) and S\ref{fS_GlassCoPt}(d) show the circularly polarized light induced magnetization dynamics with different optical helicities for two different experimental configurations. The experimental results obtained are compared in Fig. S\ref{fS_GlassCoPt}(e); almost no difference in the phase of magnetization precession is observed. The phase analysis of the magnetization precession with two experimental configurations are shown in Figs. S\ref{fS_GlassCoPt}(f) and S\ref{fS_GlassCoPt}(g), in which no sign reversal is observed.
The sign reversal of the optical helicity dependent terahertz emission was observed in Ref. \cite{Huisman2016}.
However, this result is consistent with the fact that the sign of the generated spin does not depend on the polarity of symmetry [Eq. (4) in the main text]\cite{Edelstein1997}.

